\documentclass[twocolumn,showkeys,prx,superscriptaddress,floatfix]{revtex4-2}
\usepackage[utf8]{inputenc}
\usepackage{amsmath}
\usepackage{graphicx}
\usepackage{amsfonts}
\usepackage{amssymb}
\usepackage{stmaryrd}
\usepackage[normalem]{ulem}
\usepackage{xcolor}

\usepackage[colorlinks=true, allcolors=blue]{hyperref}
\usepackage[capitalise]{cleveref}
\usepackage[english]{babel}
\usepackage{csquotes}
\usepackage{blindtext}
\usepackage[letterpaper,top=2cm,bottom=2cm,left=2cm,right=2cm,marginparwidth=1.75cm]{geometry}

\newcommand\numberthis{\addtocounter{equation}{1}\tag{\theequation}}

\usepackage{comment}

\begin{document}

\title{
How do particles with complex interactions self-assemble?
}

\author{Lara Koehler}
\affiliation{Université Paris-Saclay, CNRS, LPTMS, 91405, Orsay, France}
\affiliation{Max Planck Institute for the Physics of Complex Systems, Dresden, Germany}
\author{Pierre Ronceray}
\email{pierre.ronceray@univ-amu.fr}
\affiliation{Aix Marseille Universit\'e, CNRS, CINAM, Turing Center for Living Systems, 13288 Marseille, France}
\author{Martin Lenz}
\email{martin.lenz@universite-paris-saclay.fr}
\affiliation{Université Paris-Saclay, CNRS, LPTMS, 91405, Orsay, France}
\affiliation{PMMH, CNRS, ESPCI Paris, PSL University, Sorbonne Université, Université Paris-Cit\'e, F-75005, Paris, France}
\date{\today}
\begin{abstract}
 In living cells, proteins self-assemble into large functional structures based on specific interactions between molecularly complex patches.
Due to this complexity, protein self-assembly results from a competition between a large number of distinct interaction energies, of the order of one per pair of patches.
Current self-assembly models however typically ignore this aspect, and the principles by which it determines the large-scale structure of protein assemblies are largely unknown.
Here, we use Monte-Carlo simulations and machine learning to start to unravel these principles.
We observe that despite widespread geometrical frustration, aggregates of particles with complex interactions fall within only a few categories that often display high degrees of spatial order, including crystals, fibers, and oligomers.
We then successfully identify the most relevant aspect of the interaction complexity in predicting these outcomes, namely the particles' ability to form periodic structures.
Our results provide a first extensive characterization of the rich design space associated with identical particles with complex interactions, and could inspire engineered self-assembling nanoobjects as well as help understand the emergence of robust functional protein structures.
\end{abstract}

\maketitle

Multiple copies of a single protein often self-assemble to fulfill their biological functions~\cite{marsh2015structure}. The resulting assembly morphologies may be complexes of a few subunits, \emph{e.g.}, membrane channels, large but finite higher-order assemblies akin to viral capsids, or unlimited structures such as cytoskeletal fibers~\cite{goodsell2000structural}. The interactions between individual proteins are dictated by the amino-acids at their surface. These interact through a wide range of physical effects, including hydrophobic-hydrophilic interactions, polar and electrostatic forces as well as steric repulsions and shape complementarity~\cite{chothia1975principles, hu2000conservation,mccoy1997electrostatic, lawrence1993shape}, implying a wide range of interaction affinity and specificity~\cite{zhou2018electrostatic,empereur2019geometric, johnson2011nonspecific}. Despite the complexity of these interactions, the products of protein aggregation overwhelmingly fall into a small number of stereotypical aggregate morphologies. These include oligomers~\cite{stehle1996structure}, one-dimensional fibrillar structures~\cite{noree2010identification, dykes1979three, knowles2014amyloid}, and liquid condensates of finite~\cite{de2012casein} or unlimited three-dimensional sizes~\cite{li2012phase}. Three-dimensional crystals are also observed \textit{in vivo}~\cite{hartje2019protein, mcpherson2004introduction, lanci2012computational} and \textit{in vitro}~\cite{koopmann2012vivo}, and widely used  to crystallographically investigate protein structure.
These morphologies thus display a range of different dimensionalities and orientational order of the proteins. 

The relationship between the molecular structures of the protein surfaces that come into contact upon binding -- which we refer to as ``patches'' in the following -- and the morphology of the resulting aggregates is not well understood. It is for instance difficult to discriminate between the amino-acids that are involved in a protein-protein interaction and those that remain unbound~\cite{levy2013structural,moal2013scoring}. On a more practical level, we lack an effective framework to predict protein crystallization as a function of solvent conditions~\cite{mcpherson2004introduction}. When a well-defined aggregate morphology is obtained, it is typically sensitive to subtle changes in interprotein interactions. A single mutation can thus trigger the self-assembly of a soluble protein into fibers \textit{in vitro}~\cite{garcia2017proteins}. \textit{In vivo}, proteins found in different organisms with almost identical morphologies  may nonetheless assemble through completely different patches~\cite{hakim2009dimer,sinha2007subunit,lynch2017human}. Many proteins are thus increasingly believed to be equipped with multiple competing sticky patches which may or may not dominate the final assembly depending on potentially subtle factors. This competition may underpin the widely observed structural polymorphism in protein self-assembly~\cite{thirumalai2012role,nguyen2009invariant}. 

A popular theoretical approach to the relationship between protein interactions and the resulting self-assembly phase diagram is the use of so-called patchy particle models, where anisotropically patterned particles interact through a small set of short-range interactions~\cite{kern2003fluid}. Varying the number and the position of sticky patches on the particles influences both the orientational order of the particles locally~\cite{whitelam2012random,romano2020designing}, and the dimensionality and size of the aggregate~\cite{zhang2004self, karner2020patchiness,grunwald2014patterns}. Patchy particle models are useful for predicting the morphology resulting from the assembly of some specific proteins~\cite{khan2019temperature}. While some studies provide a partial exploration of the design space of patchy particles  \cite{evans2017physical,fusco2013crystallization, haxton2012design}, there is no systematic understanding of the relationship between the particle interactions and the aggregate morphology. 

\begin{figure}[t]
    \centering
    \includegraphics[width=\linewidth]{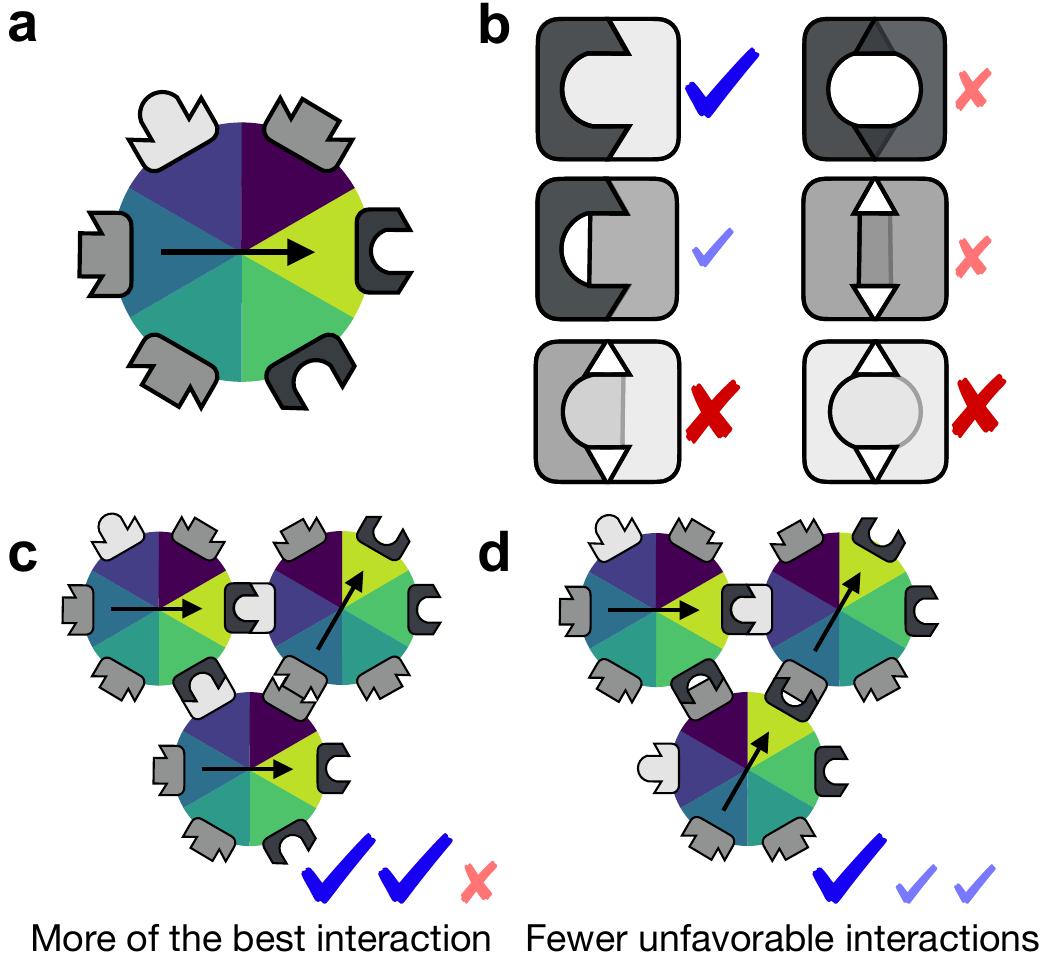}
    \caption{\textbf{Complex asymmetric interactions lead to geometric frustration.}
    (a)~Particles are asymmetrically patterned with  "lock-and-key" interacting patches. The arrow and colors indicate the particle orientation.
    (b)~Each pair of patches is governed by its own interaction strength, resulting in a large number of independent interactions. For instance, the squared-lock and the round-lock both bind to the round-key, but not with each other.
    (c)~When a particle interacts with two others through two of the best interactions, this leads to an unfavored interaction between these two neighbors.
    (d)~An alternative, possibly more favorable arrangement gives up one of the best interactions to avoid the resulting unfavored interaction. } 
    \label{fig:intro}
\end{figure}

Existing theoretical approaches to protein self-assembly leave a crucial aspect of protein interactions largely unexplored: the fact that pairs of patches have essentially independent interaction energies, due to both the variety of physico-chemical interactions involved and the combinatorial complexity of patch geometries. In this sense, their interactions are non-transitive: the fact that two patches stick to a third does not necessarily imply that they would stick to (or repel) one another. To illustrate the complex interplay resulting from the interaction between competing patches, in~\cref{fig:intro}(a) we consider a particle that is asymmetrically patterned with three types of patches whose interactions are detailed in~\cref{fig:intro}(b). The complexity of these interactions makes them pair-specific, and one cannot be deduced from the knowledge of the others: a particle with $n$ distinct patches has $\sim n^2$ independent pair interactions between patches, in contrast with simple interactions governed by a single scalar quantity such as the electrical charge, which would result in only $\sim n$ independent interactions.
Such a large set of interactions generically gives rise to a competition between competing local structures, all involving some suboptimal interactions [\cref{fig:intro}(c-d)]. Optimizing the morphology of the aggregate in the presence of this so-called frustration is a notoriously nontrivial task and usually results in polymorphism~\cite{rivier1988polymorphism,zhao2015shape,haxton2012design}.
Geometrical frustration leads to size limitation of the aggregate in other contexts, such as the self-assembly of elastically deformable particles in two and three dimensions~\cite{lenz2017geometrical,hall2022building,tyukodi2022thermodynamic}. It also influences the crystalline order in lattice particles~\cite{ronceray2019range, meiri2022cumulative}. 

In this paper, we investigate how a large number of independent interactions influences the morphologies formed by self-assembling lattice particles.  
In~\cref{sec:models}, we introduce a minimal lattice-based model with 21 independent continuous interaction parameters and show that it produces a range of equilibrium aggregate morphologies in numerical simulations. 
We then use machine learning in~\cref{sec:classification} to show that despite the complexity of the interactions, the resulting morphologies can be grouped within a small number of categories with the same aggregate dimensionality and orientational order. Particles with highly asymmetric interactions can result in nontrivial morphologies reminiscent of those found in proteins, \textit{e.g.} fibers or self-limited assemblies, and in \cref{sec:frustration} we show that such aggregates typically form as a way to avoid geometrical frustration. 
Finally, \cref{sec:learning-interpretation} presents a first foray in understanding the relationship between particle interactions and aggregate morphology by using machine learning to compare the prediction accuracy of different descriptors, each aimed to isolate specific features of our interaction parameters.

\section{Arbitrary interactions yields diverse aggregates morphologies}
\label{sec:models}
\begin{figure*}[t]
\centering
\includegraphics[ width=\linewidth]{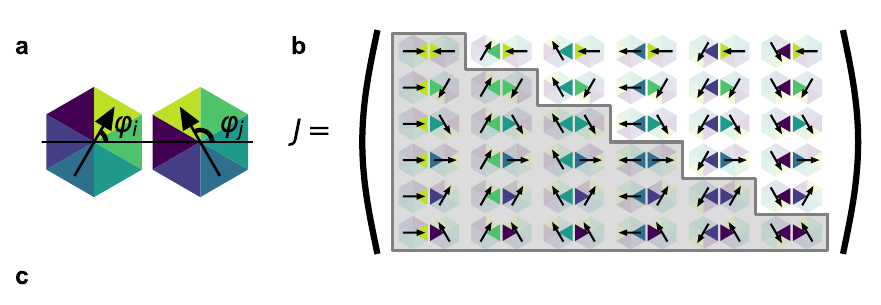}
\includegraphics[width=\linewidth]{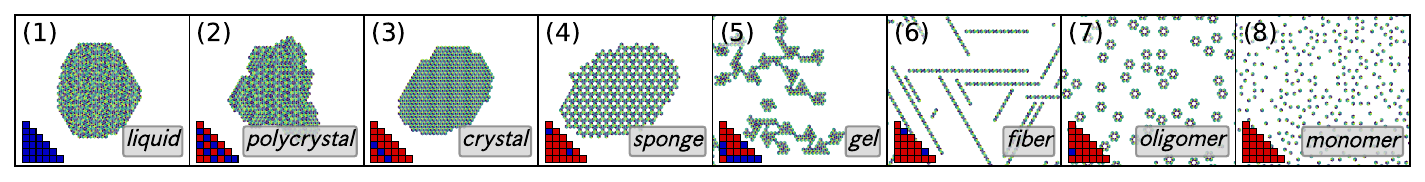}
\caption{\label{fig:introduce_Jab}
\textbf{Simple sets of local interactions lead to a large diversity of aggregates}
(a)~Lattice particles interact through their six patch-like faces. The interaction depends on the particles' relative orientations.
(b)~The set of all possible interparticle orientations can be represented in a symmetric $6\times 6$ matrix. The contents of the matrix-like interaction map $J$ can thus be summarized by specifying the lower triangular part of this matrix.
(c)~Equilibrium configurations of systems of $500$ particles for eight easily interpretable interaction maps. The lower triangular part of the interaction map $J$ is represented in the lower left corner of each panel, with blue, and red squares respectively representing interactions energies $J_{ab}= -10$, $10$ in units of $k_BT$. The labels on the bottom right of each panel indicate our nomenclature of the morphologies.}
\end{figure*}

We design a minimal model of particles with directional interactions, each of them of arbitrary strength and sign. As shown in \cref{fig:introduce_Jab}(a), we consider identical hexagonal lattice particles on a triangular lattice. Two neighboring particles who come into contact through their faces $a$ and $b$ ($a, b\in [1..6]$) experience an interaction energy $J_{ab}$. We denote the set of all interactions as $J$, and refer to it as the ``interaction map'' of the problem. Without loss of generality, we set the thermal energy $k_BT$ to one and all interaction energies between a particle and an empty site to zero (\cref{sup:invariance}), implying that $J$ fully characterizes the energetics of a system of particles. There are $6\times6$ pairs of faces, but since $J_{ab}=J_{ba}$ by symmetry, $J$ has only $21$ non-redundant elements corresponding to the interactions schematized in \cref{fig:introduce_Jab}(b). This large number of independent energies allows us to capture the large complexity illustrated in \cref{fig:intro} without reference to the microscopic physical origin of each interaction.

To characterize the influence of $J$ on particle self-assembly, we look for the equilibrium state of low-density systems in the canonical (NVT) ensemble using Monte-Carlo simulated annealing. We implement single particle and cluster Monte-Carlo moves to reach the equilibrium configuration at a modest computational cost (see~\cref{sup:Monte-Carlo}, for details). Our procedure does not mean to model a realistic dynamic for self-assembly.  We verify that the simulation results do not depend on the chosen annealing duration and particles' density (\cref{sup:equilibration}).

We first illustrate the range of possible outcomes of the simulations in~\cref{fig:introduce_Jab}(c) by choosing eight stereotypical interaction maps, which we pictorially depict in the corner of each panel. In panel~(1), a fully isotropic and attractive interaction map produces a compact aggregate devoid of orientational order akin to a liquid droplet. By contrast, panel~(2) and panel~(3) present highly anisotropic, attractive interaction maps that promote the short- or long-range orientational ordering of the particles. They thus respectively produce a polycrystal and a ordered crystal. Interaction map~(4) also promotes a small number of particle contacts, but unlike in the previous examples these can only be realized in particles with different orientations, resulting in the formation of a sponge-like morphology. Panel~(5) displays particle with competing interactions: one face of a particle can bind to several others, as evidenced by the presence of several blue entries for the same line in the interaction map. These leads to the formation of a gel, where smaller clusters of a few particles are connected to each other while the whole structure retains a large interface with the solvent. If a particle only has sticky patches located opposite each other, it forms a fiber, as in panel~(6). 
Finally, a single attractive interaction favoring misaligned particles yields hexamers in panel~(7), and the absence of interactions in panel~(8) results in a gas. These categories recapitulate many morphologies observed in aggregates of proteins or patchy particles, thus outlining the ability of complex interactions to induce complex aggregates even in our comparatively simple lattice-based model. 

\begin{figure}[b]
    \includegraphics[width=01\linewidth]{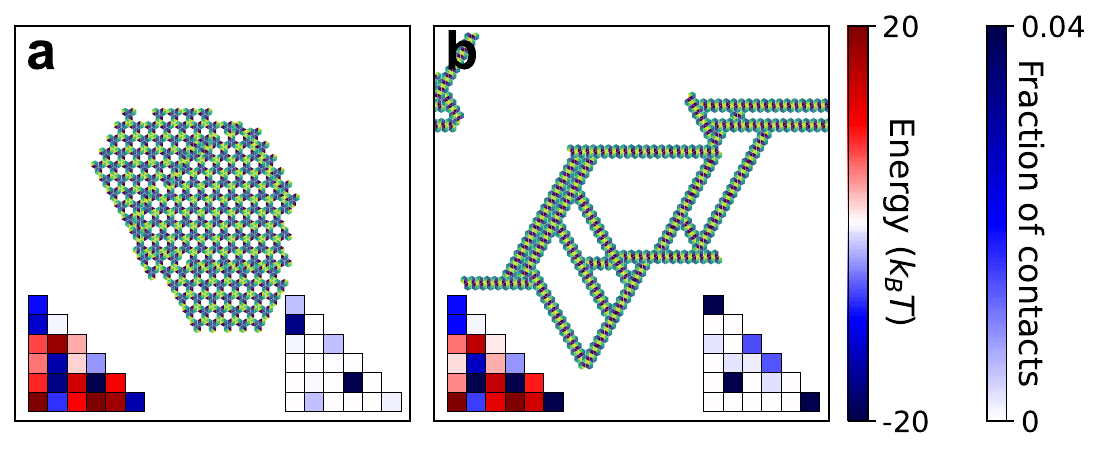}
    \caption{\textbf{Similar interaction maps can lead to very different aggregate morphologies}. 
    The two interaction maps of panels~(a) and (b) have the same favored and unfavored interactions, yet result in very different final morphologies (a sponge and a fiber per our nomenclature).}  
    \label{fig:sameLELdiffCat}
\end{figure}

Beyond these simple examples, we use arbitrary interaction maps to show that our model qualitatively recapitulates the effects of the competition between sticky patches in proteins. In \cref{fig:sameLELdiffCat}, we thus show two almost-identical interaction maps leading to very different aggregate morphologies depending on which one of two competing sets of interparticle contacts is more stable than the other. To highlight the differences between these two sets, in each panel we display the contact map $\langle N\rangle$, which gives the overall frequency of each pair of particles in the simulation. While the most frequent contacts correspond to favorable contacts in the interaction map, not all favored interactions are observed. This demonstrates that having many arbitrary, continuous interaction values leads to a relationship between interaction map and contact map that is both highly sensitive and nontrivial.

\section{Aggregate morphologies fall within a few stereotypical categories}
\label{sec:classification}
As aggregate morphologies sensitively depend on the exact values of the underlying interactions, one may wonder whether new aggregate categories beyond the eight pictured in \cref{fig:introduce_Jab}(c) could emerge for a fully general interaction map. To begin to answer this question, here we simulate a large number of randomly chosen interactions maps and classify the resulting morphologies with the help of a machine learning algorithm.

\begin{figure*}[t]
\centering
\includegraphics[width=\linewidth]{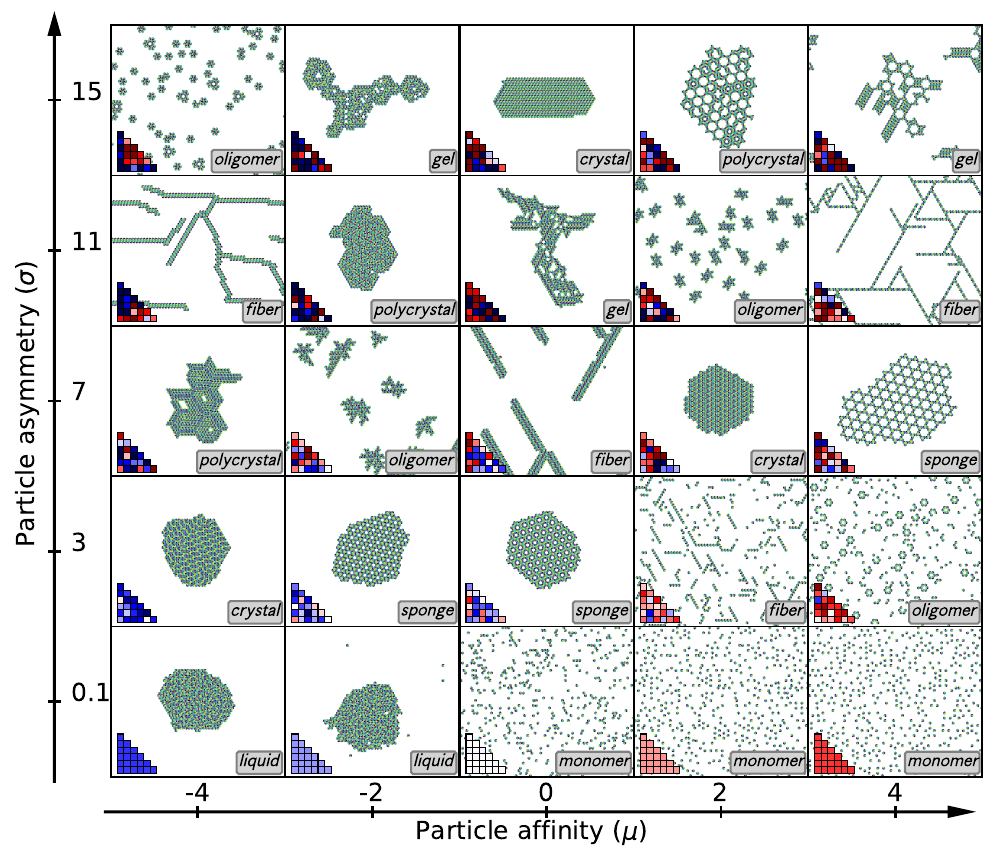}
\caption{\label{fig:random_examples} \textbf{Random interactions reproduce the aggregate diversity observed in simple interaction maps.} For each value of affinity and asymmetry, we show a randomly drawn interaction map (bottom left matrix, with the color scale of~\cref{fig:sameLELdiffCat}), a snapshot of the result of the simulated annealing, and the aggregate category, as in~\cref{fig:introduce_Jab}. Here and in the following, the lattice has $60\times60$ sites, and there are $500$ particles. We verify in \cref{fig:Nparticles_dependence} that the aggregate morphologies are unchanged when considering 10 times more particles at similar density.}
\end{figure*}

To guide our exploration, we reason that two major determinants of a particle's self-assembly behavior are its affinity, \emph{i.e.}, its average propensity to stick to other identical particles, and its asymmetry, \emph{i.e.}, its deviation from an isotropic interaction profile. Experimentally, the former can in principle be tuned independently of the latter through, \emph{e.g.}, depletion interactions. We thus choose to respectively model the affinity and asymmetry using two independent parameters $\mu$ and $\sigma$. We draw each of the 21 independent parameters of our interaction map independently of the others from the following Gaussian distribution:
\begin{equation}\label{eq:proba}
    P(J_{ab})=\frac{1}{\sqrt{2\pi\sigma^2}}\exp\left[-\frac{(J_{ab}-\mu)^2}{2\sigma^2}\right].
\end{equation}
We show typical aggregates resulting from several $(\mu,\sigma)$ values in \cref{fig:random_examples}. At low asymmetry $\sigma$, liquids or monomers dominate depending on the affinity $\mu$, consistent with the absence of orientational preference of the particles. Larger values of the asymmetry yield diverse morphologies. Despite some variability -- \emph{e.g.}, the varying widths and branching rates of the fibers in \cref{fig:random_examples} -- all aggregate morphologies fall within the categories enumerated in~\cref{fig:introduce_Jab}(c) (see labels on each image).

\begin{figure}
    \centering
    \includegraphics[width=1\linewidth]{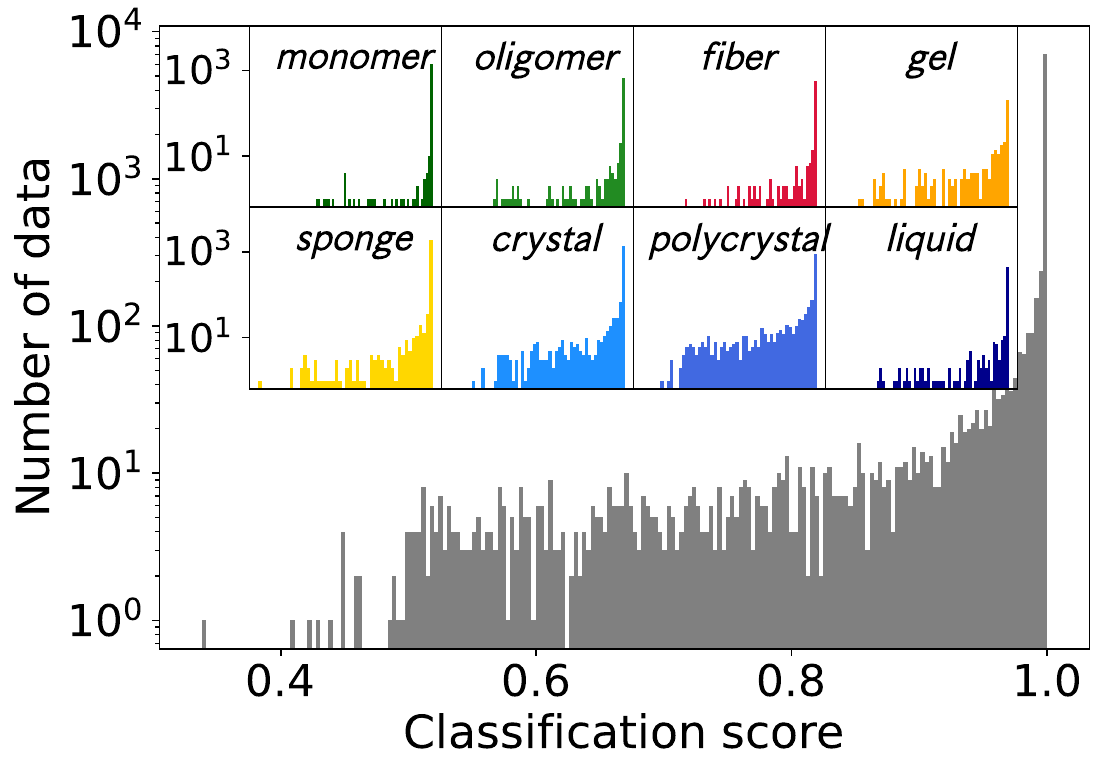}
    \caption{\textbf{We classify the aggregates without ambiguity with machine learning.} We show, in log-scale, the histogram of the predictions scores of the neural-network for the whole dataset (gray), and for the data classified in each category (in the insets, the axes are the same as the main figure). Most of the prediction scores (the probabilities to belong to the predicted category) are close to unity, suggesting an unambiguous classification.}
    \label{fig:highest_scores}
\end{figure}

We make our morphological classification more systematic by generating a large number of interaction maps and studying the resulting aggregates. We choose 45 different $(\mu,\sigma)$ couples corresponding to $5$ values of affinity and $9$ values of asymmetry: $\mu \in \{-4, -2, 0, 2, 4\}$, $\sigma \in \{0.1, 1, 3, 5, 7, 9, 11, 13, 15\}$. For each of these, we generate $200$ interaction maps and run the numerical annealing procedure described above. 
We characterize each morphology by computing a few geometric properties, namely the average aggregate size, porosity, and surface to volume ratio. Out of the resulting $9000$ aggregate morphologies, we manually classify $738$ randomly chosen ones within our eight categories. 
As is apparent from Supplementary~\cref{fig:label_monomer,fig:label_oligomer,fig:label_gel,fig:label_fiber,fig:label_sponge,fig:label_crystal,fig:label_polycrystals,fig:label_liquid}, we do not find the need for any new category during this process. We then use this manually labeled set of categorical data to train a simple feedforward neural network to predict the label of a given aggregate morphology using the corresponding interaction map, contact map, and aforementioned geometric properties as descriptors. For any set of descriptors, the network outputs a set of eight scores that sum to one, each assigned to a category. We choose the category with the highest score as the classifier's prediction. This procedure yields reliable results, with $99.7\%$ correct prediction on the training set and $99.3\%$ correct predictions on the test set (see~\cref{sup:machine_learning} for details on the method and the classification). For instance, we show in  \cref{subsec:entropy_local_structures} that the network successfully segregates large aggregates with different levels of orientational order between the sponge, crystal, polycrystal and liquid categories.  We use the network to classify our whole dataset of 9000 morphologies,  and evaluate the quality of each prediction from the value of the highest probability, which we refer to as the \textit{score} of the prediction. An ideal, unambiguous classifier should give scores close to unity, much larger than the $\sim 1/8^{\mathrm{th}}$ probability associated with a random classification of categories of approximately equal sizes. \cref{fig:highest_scores} shows the histogram of the scores for our classifier with a vertical logarithmic scale. For each category, a large majority of the scores are close to unity, with only $8.2\%$ of the morphologies having a score below $0.9$ ($3.3\%$ below $0.7$). Low scores typically occur in systems where two morphologies are present simultaneously, as shown in Supplementary~\cref{fig:low_scores}. This successful outcome confirms both that our eight categories are sufficient to classify our sample of morphologies without significant ambiguities, and that the category that an aggregate belongs to can be determined by specifying the particle interactions alongside a few geometrical characteristics.

\begin{figure}[t]
\centering
\includegraphics[width=\linewidth]{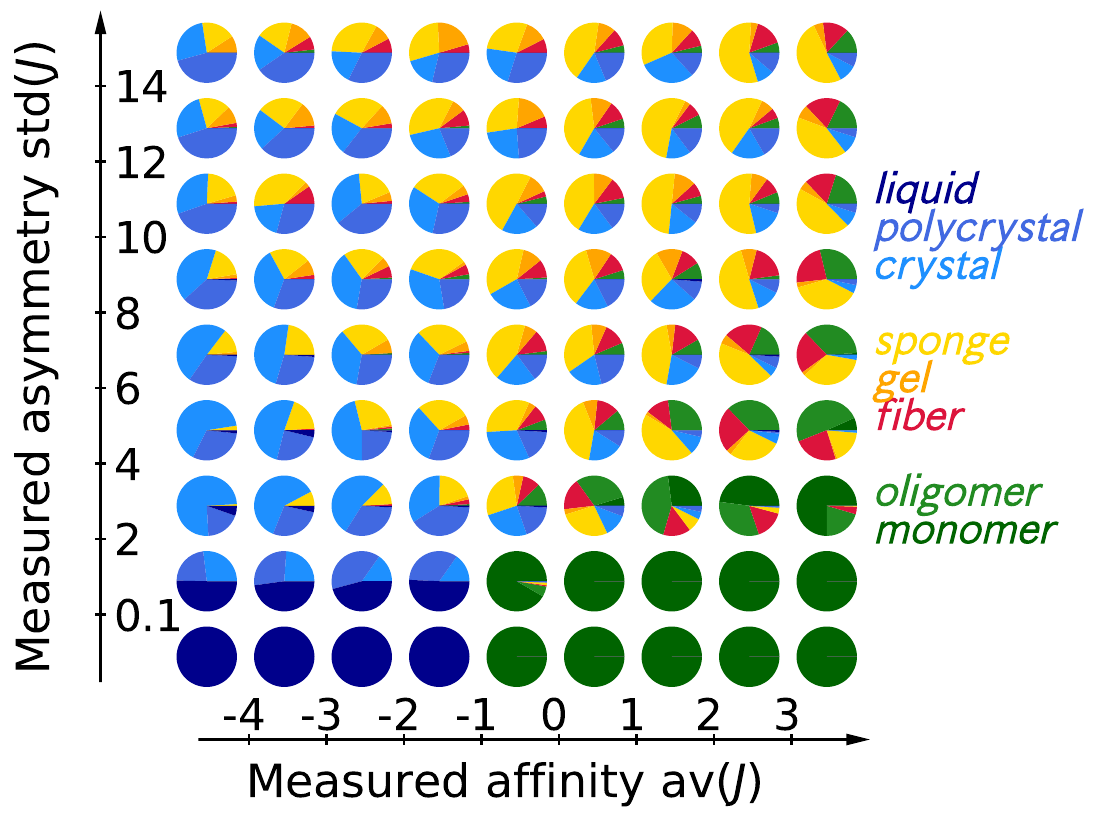}
\caption{\label{fig:phase_diagram} \textbf{Asymmetric interactions frequently lead to aggregates of reduced sizes or dimensionality.} Each pie-chart shows the statistics of aggregate categories formed from random interaction maps as a function of affinity, and anisotropy. The $9000$ randomly drawn interaction maps are binned according to the measured affinity, and anisotropy, with precision of $1k_BT$. 
Each pie-chart summarizes the information of 50 to 289 interaction maps. Aggregate categories are determined by supervised machine learning. The same graph with the data binned according to the affinity and anisotropy of the distribution is shown in \cref{fig:phase_diagram_drawn_binning}}
\end{figure}

By applying our classifier to all the unlabeled aggregates among our 9000 interaction maps, we conduct an extensive statistical analysis of the influence of the affinity $\mu$ and the asymmetry $\sigma$ on the aggregate morphology. We present our results in \cref{fig:phase_diagram}, which confirms the tendencies identified in \cref{fig:random_examples}. Very sticky particles thus favor the formation of infinite aggregates (liquid, crystal and sponges, in blue and on the left of the diagram), while repulsive, high-symmetry particles form monomers (the bottom-right of the diagram is mostly light green). By contrast, nontrivial aggregates form mostly for particles that are on average repulsive, and highly asymmetric (upper right region of the diagram). A more specific characterization however appears difficult from this data alone, as most values of $(\mu,\sigma)$ within this region produce very diverse collections of morphologies.

\section{Particles form slender, small or porous aggregates to avoid geometrical frustration}
\label{sec:frustration}
To further elucidate the relationship between interactions and aggregate morphology within the nontrivial repulsive/asymmetric region of \cref{fig:phase_diagram}, we reason that geometrical frustration (\cref{fig:intro}) should penalize the formation of compact crystals and liquid aggregates. According to this reasoning, particles that display many incompatible interactions should tend to form aggregates of lower size, of lower dimensionality, or with higher porosity.

\begin{figure}[t]
    \centering
    \includegraphics[width=1\linewidth]{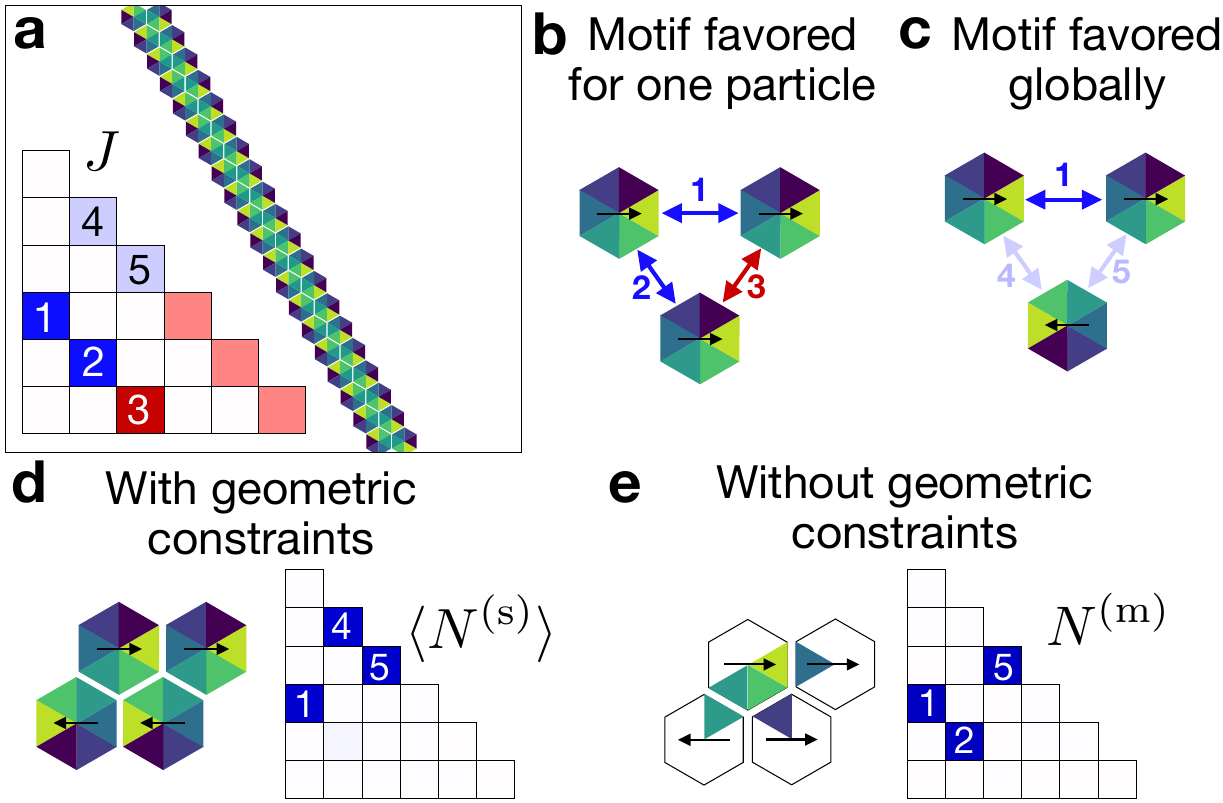}
    \caption{\textbf{We quantify geometrical frustration as the energy surplus associated with the geometric constraints}. (a) Interaction map $J$ leads to a fiber (same color codes as~\cref{fig:sameLELdiffCat}). 
    (b)~When a single particle (here on the top left) realizes the most favorable interactions with its neighbors, it leads to an unfavorable interaction between these neighbors (labeled ``3'' here).
    (c)~Minimizing the energy of the three particles together leads to the same fiber motif observed in panel~(a).
    (d)~In the simulation, we measure the contact map $\langle N^{\mathrm{(s)}}\rangle$ that takes into account the geometric constraints associated with such local particle arrangements.
    (e)~By contrast, in our constraint-free minimization we determine the contact maps that minimizes the energy $N^{\mathrm{(m)}}$ without constraints of this type. We quantify frustration as the relative energy difference between the protocols illustrated in these last two panels ($f=0.75$ in that example).} 
    \label{fig:frustration_explain}
\end{figure}

We define a quantitative measure of frustration whose design is illustrated in \cref{fig:frustration_explain}. The interaction map of panel~(a) implies a competition between two local structures. In the first structure, shown in panel~(b), the geometry of the particles imposes a larger-scale geometrical constraint. Specifically, it imposes that two favorable interactions can only be obtained at the cost of an unfavorable one. As a result the motif of panel~(c), which only comprises favorable interactions, albeit weaker ones, is favored overall. We propose that the amount of geometrical frustration associated with an interaction map can be quantified as the amount of favorable interaction energy that is ``lost'' due to the aforementioned geometrical constraints. This quantity can be measured by comparing the equilibrium energy of a numerical simulation, where these constraints are present [panel (d)], to a situation where these constraints are removed [panel (e)]. To engineer such a situation, we imagine a mean-field system where each particle is broken down into its six constitutive faces, and where all faces are free to associate in pairs irrespective of their provenance (see \cref{sup:gas_faces} for the calculation of $N^{(m)}$). As a result, the formation of the two most favorable interactions no longer forces an unfavorable interaction, as illustrated in panel~(e).

Operationally, for a given interaction map $J$, we respectively denote by $\langle N^{\mathrm{(s)}}\rangle$ and $ N^\mathrm{(m)} $ the average contact map obtained from our simulations and the contact map obtained from this new constraint-less free-energy minimization. We then define our measure of the relative frustration as
\begin{equation}
    f =  \frac{\left\langle E^{\mathrm{(s)}}\right\rangle - E^{\mathrm{(m)}}}{\left\langle E^{\mathrm{(s)}}\right\rangle}, 
    \label{eq:relative_frustration}
\end{equation}
where the average simulated energy reads
\begin{equation}
    \left\langle E^{\mathrm{(s)}}\right\rangle =  \sum_{a\leq b} \left\langle N_{ab}^{\mathrm{(s)}}\right\rangle J_{ab} 
\end{equation}
and similarly for $E^{\mathrm{(m)}}$.

Our example of \cref{fig:frustration_explain} illustrates the putative effect of frustration on the aggregate morphology. Because the configuration of panel~(b), where all particles are aligned, is ruled out by frustration, the system tends to select the more complex configuration of panel~(c). This local configuration could in principle lead to a dense aggregate, \emph{e.g.}, a crystal of alternating particles. In practice, however, this would require forming additional contacts besides those represented in panel (c), and those contacts are penalized by weak repulsive interactions that appear as light red squares in \cref{fig:frustration_explain}(a). This constitutes a new source of frustration for any hypothetical dense aggregate. We thus predict that densely packing such particles in a simulations box without any empty sites would result in a fairly large frustration $f$. By contrast, the dilute system of panel~(a) avoids all unfavorable interactions by forming a fiber, resulting in a lower value of $f$. More generally, we speculate that an effective way for a dilute system to avoid unfavorable interactions is to incorporate empty sites in its morphology. The incentive to do so should be larger in interaction maps that result in a large ``dense frustration'' $f_\text{dense}$, implying that this dense frustration could be correlated with the final aggregate morphology.

\begin{figure*}
    \centering
    \includegraphics[width=\linewidth]{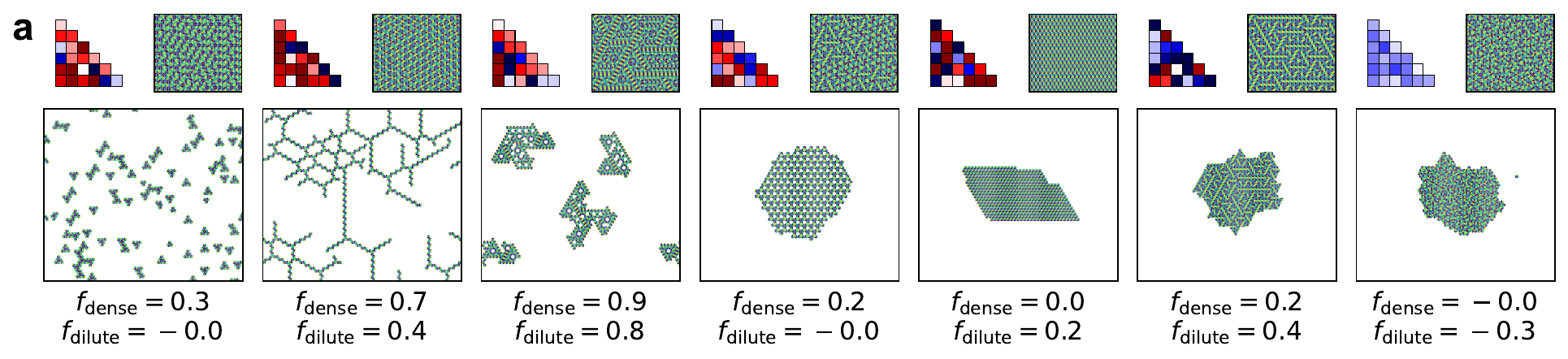}
    \includegraphics[width=\linewidth]{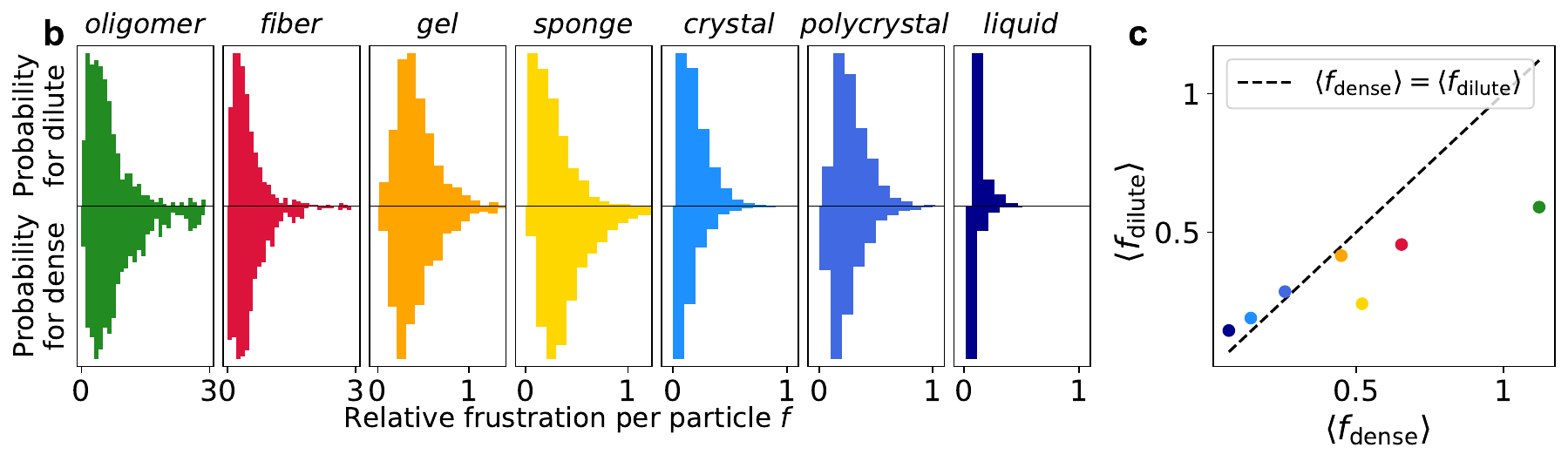}
    \caption{\textbf{Particles avoid frustration by self-assembling into non-compact aggregates.}
    In a dense system, individual particles have no choice but to interact with their neighbors, resulting in an overall higher frustration than in dilute systems.
    (a)~Example of dense and dilute equilibrium configurations and corresponding relative frustrations as defined in~\cref{eq:relative_frustration} for seven representative interaction maps. The last two examples are exceptions to the average trend of higher dense frustration. Dense configuration are simulated for $484$ particles in a lattice of $22\times22$ sites.
    (b)~Histogram of the probability density of relative frustration for each of our seven categories of aggregating (non-monomer) systems.
    (c)~Corresponding averages.}
    
    \label{fig:stat_frustration}
\end{figure*}

To test this correlation and the idea that the formation of small, slender or porous aggregates leads to a reduction in frustration, we compute $f_\text{dense}$ and $f_\text{dilute}$ for each interaction map in our sample of $9000$, according to \cref{eq:frustration}. Here, $f_\text{dense}$ is calculated from simulation results of systems with the same interaction map, number of particles, but density $1$. The examples outlined in \cref{fig:stat_frustration}(a) validate our speculations: the five interaction maps on the left display a high dense frustration, which they relax in a dilute setting by taking advantage of empty sites. By contrast, the two rightmost particles display low levels of dense frustration. When diluted, they form compact aggregates with an internal organization resembling the dense systems. By contrast with the first group of five, in these systems the boundaries of the dilute aggregate are less energetically favorable than the bulk, implying a dilute frustration higher than its dense counterpart. These trends are confirmed statistically in the histograms and averages of \cref{fig:stat_frustration}(b) and (c). We thus conclude that most of our systems are frustrated, and that a high dense frustration is associated with a non-compact dilute morphology that enables a reduction of the frustration.

\section{The ability to assemble into periodic motifs predicts the aggregate category}
\label{sec:learning-interpretation}
While our findings on the role of frustration provide insights into the physics that underpins the self-assembly of particles with complex interactions, the histograms of \cref{fig:stat_frustration}(b) overlap too much for the number $f$ to serve as a reliable predictor of the resulting aggregation category. To better understand the most crucial aspects of the interaction map, here we again train neural networks to predict the outcome of self-assembly, but this time while intentionally providing them only with partial information.

\begin{figure}
\centering
\includegraphics[ width=\linewidth]{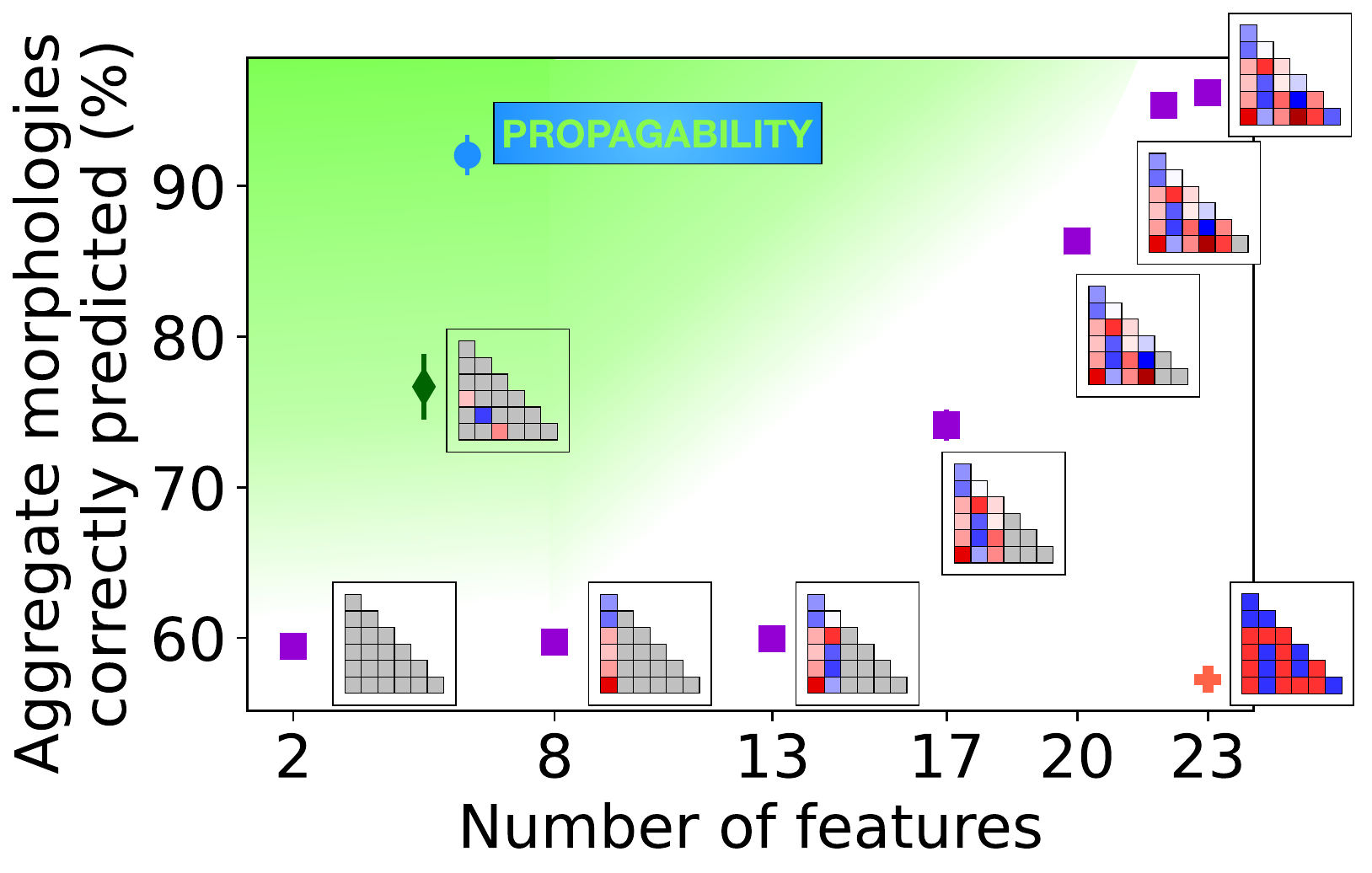}
\caption{\label{fig:interpretationML} \textbf{Propagability, a measure of the ability to form periodic motifs, is a very successful predictor of aggregate morphology.}
\emph{Purple squares:} A neural network can accurately predict the aggregation category when provided with the full aggregation map (rightmost purple square), although this ability rapidly degrades when masking some of the features of this map (grayed out in the inset).
\emph{Green gradient:} Strategies that result in an accuracy above the purple symbols outperform this crude baseline. This category includes the specification of the line interactions (green diamond symbol) and propagability (blue circle). By contrast, the sole sign of each interaction energy is a very poor predictor of the morphology (orange cross).
} 
\end{figure}

From a formal point of view, our Monte Carlo simulation outputs the aggregation category corresponding to an interaction map from the specification of its $21$ independent components. We first verify that a neural network can emulate this computation given a large enough training set to learn the full high-dimensional phase diagram. More specifically, we perform a closely related test on our sample of 9000 by providing a neural network with the interaction map, as well as the average av($J$) and standard deviation std($J$) of the energies of interaction map $J$. The quantities av($J$) and std($J$) are closely related to $\mu$ and $\sigma$, with the difference that the former relate to an individual $J$ and the latter to the underlying probability distribution [\cref{eq:proba}].
As shown in \cref{fig:interpretationML}, the resulting prediction accuracy is close to 100\%.
Compared to this ideal case, a neural-network prediction based on fewer than $23$ scalar values --~referred to as ``features'' in the following~-- should be less accurate as it proceeds from a more limited amount of information. To get a sense of the expected decrease in accuracy upon a decrease in the number of features, we train neural networks based on a restricted amount of information by omitting to provide them with some of the components of the interaction map (see details in \cref{subse:method_interpretation}). As shown in \cref{fig:interpretationML}, under this protocol the predictive power of the neural networks decreases monotonically as the number of features decreases. In the most extreme case, the accuracy falls to less than 60\%, when only the measured $\mathrm{av}(J)$ and $\mathrm{std}(J)$ are provided. 

While our wholesale masking of the interaction map 
sets our expectation for the accuracy expected from a given number of features, we reason that some better-chosen descriptors could outperform this baseline. Here we look for such descriptors as a means to identify the aspects of the interaction maps most relevant to the outcome of self-assembly. In a first test of this idea, we ask whether the aggregate morphology could simply be determined by the specification of which of its interactions are attractive \emph{vs.} repulsive, irrespective of their intensity. Such an outcome would be contrary to our previous discussions of the role of frustration in our system, whereby a subtle balance between the magnitude of several interactions determines which local structures are actually chosen by the system. We see in \cref{fig:interpretationML} that a neural network provided solely with $\mathrm{av}(J)$, $\mathrm{std}(J)$ and the signs of the individual interactions (not their magnitudes), performs almost as poorly as one that only has access to $\mathrm{av}(J)$ and $\mathrm{std}(J)$. This finding thus further strengthens our conclusion that frustration plays an important role here.

In a second approach, we reason that some components of the interaction map are more conducive than others to the formation of aggregates of large sizes. Interactions that promote identical orientations between neighboring particles may thus favor crystals, as in example (2) of \cref{fig:introduce_Jab}(c). By contrast, if only one of these three interactions is favorable, fibers tend to form. We refer these as ``line interactions'', and attempt to predict the aggregation category from their three values alongside $\mathrm{av}(J)$ and $\mathrm{std}(J)$. As shown in \cref{fig:interpretationML}, this procedure far outperforms the $5$-features baseline. We interpret this success by noting that line interactions enable the formation of periodic aggregates, and that an enhanced ability to form such aggregates in one or two directions is a strong predictor of the formation of fibers and crystals.

\begin{figure*}
    \centering
    \includegraphics[width=\linewidth]{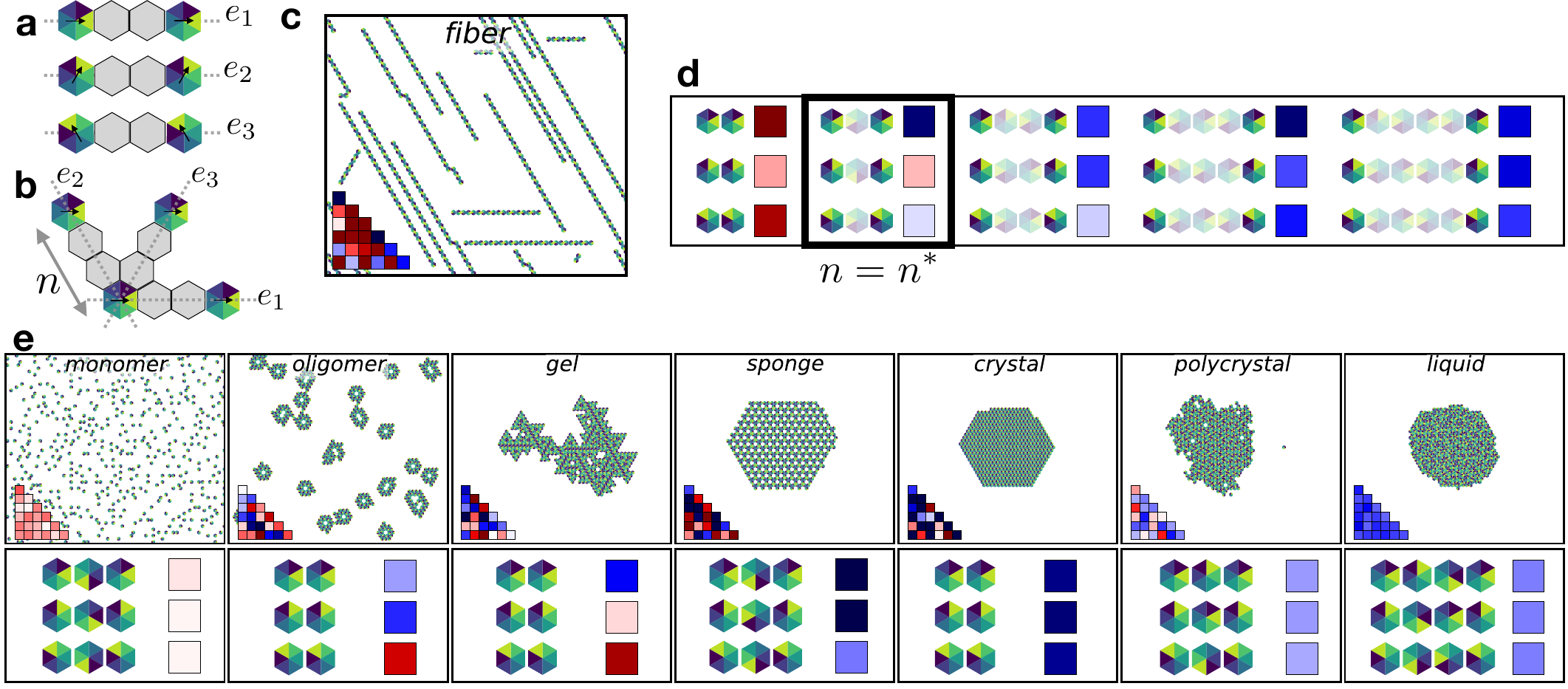}
    \caption{
    \textbf{We define an interaction map's propagability as a measure of the energy of its best periodic motifs.}
    (a)~We assess linear periodic motifs by enumerating the orientations of the gray particles and choosing the combination of orientations with the lowest energy.
    (b)~When put in the context of our triangular lattice, the three motifs of the previous panel characterize all three main lattice directions.
    (c)~An example interaction map.
    (d)~All $n\leq 7$ periodic motifs associated with this interaction map, with the associated energies per particle indicated as colored squares.
    The period of the best motif ($n=n^*=2$) is highlighted. In this example, only one of the three $n=n^*$ motifs is energetically favorable, leading to a fiber morphology.
    (e)~Other examples of $n=n^*$ motifs and associated energies.
    }
    \label{fig:propag_example}
\end{figure*}

To further exploit this insight, we note that the specification of line interactions only captures the ability to form periodic structures with a period equal to one while many of our aggregates display higher-order periodic structures. We design a more suitable predictor inspired by example (4) of~\cref{fig:introduce_Jab}(c). In this example, a fiber emerges from a combination of two interactions where particles are anti-aligned, resulting in a periodicity of $2$ lattice sites. This suggests that in this case, the nearest-neighbor line interaction discussed above can usefully be replaced by an effective second-nearest-neighbor interaction mediated by an anti-aligned particle. As illustrated in \cref{fig:propag_example}(a), we analogously define effective $n$th-neighbor interactions by filling the $n-1$ sites joining two identically-oriented particles with particles in the most energetically favorable orientations. Here we disregard particles outside of the straight line joining the particles, and consider all three possible orientations of the identically oriented particles. As shown in \cref{fig:propag_example}(b), positioning the three $n$th neighbor interactions on our triangular lattice mimics the configuration encountered in our original definition of the line interactions,  albeit with a larger mesh size. We illustrate our procedure for the interaction map of \cref{fig:propag_example}(c), for which all effective interactions for $n$ ranging from 1 to 5 are displayed in \cref{fig:propag_example}(d). We find that the best periodic motif arises for a value $n=n^*$, which in this case equals 2. This indicates that the system's best chance at forming a periodic aggregate is for a period 2. By further examining the direction-specific period-2 energies $e_1^*$, $e_2^*$, $e_3^*$ displayed as colors in \cref{fig:propag_example}(d), we find that only one of them is very favorable. This suggests that fibers should be most favorable among the period-2 structures, consistent with the result of the Monte-Carlo simulation of \cref{fig:propag_example}(c). The examples of \cref{fig:propag_example}(e) further indicate that a larger number of favorable $n=n^*$ motifs is associated with denser aggregates. This suggests that the specification of the energy of these motifs could be indicative of the final morphology of the aggregate.

To put this intuition on a more quantitative basis, for each of our interaction maps we compute the vector $(\mathrm{av}(J), \mathrm{std}(J), e_1^*, e_2^*, e_3^*, n^*)$, or ``propagability'' of the interaction map, and assess its power as a 6-features predictor of the aggregate morphology (\cref{subsec:propagability}). We find that its $92\%$ success rate far outperforms our other attempts, which we all describe in \cref{subsec:other_interpretation}. By and large, these alternative attempts are based on averages of several interactions. They are thus presumably less successful than the propagability at capturing details of particles' preference for certain local organizational, as well as their ability to tile the plane, both of which are tied to the presence of geometrical frustration. The second panel of Fig. 10e illustrates a nontrivial effect of the competition between several interactions and the usefulness of av($J$) and std($J$) in the propagability. In this example, only one of the three periodic motifs probed by the propagability is favorable, implying that particles tend to assemble into fibers. However, a relatively large std($J$) implies the likely existence of many additional attractive interactions. These interactions cause new fibers to form off the side of existing fibers, and result in the formation of oligomers that are essentially clumps of short sticky fibers. This morphology is indeed correctly predicted by our propagability-based neural network. More broadly, we show in Appendix~\cref{fig:interpretationML_supp} that the predictive power of the propagability decreases significantly when we remove av($J$) and std($J$) from it, although it still remains very high compared to the other predictors we investigated. We thus conclude that the ability of the particles to form a structure that can propagate in one of several lattice directions determines the aggregate morphology.

\section*{Discussion}
The model introduced here reveals the effect of a type of complexity that is largely disregarded in existing self-assembly models, namely non-transitive, pair-specific, highly asymmetric interactions. Despite the enormity of the associated parameter space, we find that they produce only a few stereotypical morphologies reminiscent of those encountered in protein aggregates. This suggests that the frustrated self-assembly of complex particles may be dominated by a few universality classes, whereby few of the details of the local interactions between particles are relevant to understanding the resulting large-scale morphologies.

This interpretation is supported by our ability to predict these morphologies from our ``propagability'', \emph{i.e.}, a coarse-grained version of the interactions between neighboring particles. While revealing of the mechanisms at work within the examples presented here, this descriptor leaves out several important features, including the role of particles lying in the ``holes'' between the straight gray lines of \cref{fig:propag_example}(b). It nevertheless indicates that a more systematic renormalization group approach could allow us to go beyond qualitative statements and quantitatively identify which features of the interaction map are most relevant to the aggregates' large-scale morphology. Our lattice model offers an ideal setting for such approaches. It is indeed amenable to decimation techniques developed in the early days of the study of critical phenomena \cite{maris1978teaching}, unlike existing models for self-assembly in the presence of frustration, which typically feature particles with continuous translational degrees of freedom. The simple lattice model introduced in this paper thus provides a generic and rigorous framework to determine universal features of self-assembly in future studies.

The universality classes discussed here could provide a major step in unifying observations of common features in many disparate models of frustrated self-assembly. Frustration has indeed traditionally been attributed to the presence of particles with ill-fitting shapes~\cite{sadoc1999geometrical}, or to the presence of incompatible interactions. A simple example of the latter is the simple case of the antiferromagnetic Ising model~\cite{wannier1950antiferromagnetism}, and more recent studies have also considered continuous order parameters \cite{meiri2022bridging, meiri2022cumulative}.
Such effects have traditionally been studied in dense media, where frustration may strongly influence the local organization of the system, but tends to vanish upon repeated renormalization~\cite{ronceray2019range}.
By contrast, in the context of self-assembling dilute particles, frustration influences the shape of the boundary of the aggregate, and may thus remain relevant on large scales. This leads to fibrous objects and morphologies with internal holes in a wide range of settings, ranging from particles with a frustrated internal degree of freedom to colloidal self-assembly on a curved surface \cite{lenz2017geometrical,hackney2023dispersed, xia2011self,sciortino2004equilibrium, schneider2005shapes, meng2014elastic}.
Qualitatively, such morphologies are well explained by the frustration avoidance mechanism discussed in the present work and illustrated in \cref{fig:stat_frustration}. However, no common language has yet emerged to quantitatively describe the associated structure selection mechanisms independent of the details of each model.

Such a robust physical framework could help predict the outcome of protein self-assembly. Indeed, determining protein-protein interfaces and oligomer shapes of unknown proteins remains difficult for proteins for which detailed structural information is not available in the Protein Data Bank (PDB)~\cite{wicky2022hallucinating}. So far, estimates of the binding energies of protein contacts are primarily performed by measuring how often these contacts are observed in the PDB~\cite{lensink2019blind}. Our results, however, emphasize that pair interactions that are not observed are not necessarily unfavorable. Instead, geometrical frustration leads to a nontrivial relation between the interaction map and the contact map. 
Moreover, in living cells transitions from one protein aggregate morphology to another occur following changes related to individual binding sites (for example, through phosphorylation or binding to a ligand \cite{nishi2011phosphorylation,hayouka2007inhibiting}), or to a global shift in the binding (free) energies (\emph{e.g.}, through a change in temperature~\cite{bousset2003native}).

Those modifications of the binding energies and their influence on the aggregate morphology are a typical example of the type of complexity formalized for the first time by our model. Additionally, three-dimensional extensions thereof would display an even larger level of such complexity due to additional sources of frustration due to twisting and chiral effects as well as the presence of more numerous independent interactions, \emph{e.g.}, $84$ for cubic particles.

Self-assembly is a valuable tool to build complex materials on small scales, for instance using  colloids, proteins or DNA-based subunits whose interactions can be tailored to a very large extent \cite{ding2009fabrication,zhu2021protein,zhang2014dna}. 
The diversity of aggregate morphologies observed here could inspire such designs, from self-limited oligomers to fibers with widths larger than the size of one particle. 
This mechanism of self-limitation for colloidal self-assembly has not been previously reported~\cite{hagan2021equilibrium}, and could be used as a design strategy for, \emph{e.g.}, DNA origami~\cite{berengut2020self}.
We also observe porous materials, a category with useful storage and mechanical properties~\cite{jurvcek2021hexagonal}. 
We do not observe aggregates of fractal dimensions \cite{lomander2005hierarchical} or quasi-crystals \cite{engel2015computational}, which is not an indication that such aggregates could not be observed with particles of complex interactions -- rather, these morphologies are intrinsically repressed in lattice models. 
Overall, we suggest that self-assembly based on a collection of many identical particles with highly asymmetric interactions could provide a more robust alternative to traditional designs based on multiple constituents, in which even very small non-specific interactions can be very detrimental to the self-assembly yield~\cite{murugan2015undesired}.

\acknowledgements{M.L. was supported by Marie Curie Integration Grant PCIG12-GA-2012-334053, “Investissements d’Avenir” LabEx PALM (ANR-10-LABX- 0039-PALM), ANR grants ANR-15-CE13-0004-03, ANR-21-CE11-0004-02 and ANR-22-CE30-0024, ERC Starting Grant 677532 and the Impulscience program of Fondation Bettencourt-Schueller. M.L.’s group belongs to the CNRS consortium AQV.  P.R. was supported by France 2030, the French National Research Agency (ANR-16-CONV-0001) and the Excellence Initiative of Aix-Marseille University - A*MIDEX. L.K. was supported by Ecole nationale des ponts et chaussées.}

\onecolumngrid
\appendix

\section{Methods}
\label{sec:mat_methods}

\subsection{Monte-Carlo simulation}
\label{sup:Monte-Carlo}

We determine the equilibrium configuration of particles with a given interaction map with Monte-Carlo Metropolis-Hastings simulated annealing coded in C++. Here, we explain the Monte-Carlo steps and annealing protocol. The justification for the parameter choices are given in~\cref{sup:equilibration}. Most of the methods described in \cref{sec:mat_methods} and \cref{sec:SI} were also described in \cite{koehler2023principles}.

A fixed number of particles is placed on a two-dimensional triangular lattice with periodic boundary conditions. Throughout the study, we choose a lattice of $N_\mathrm{sites}=30\times 30$ lattice sites, and $100$ particles (except in the systems shown in~\cref{fig:introduce_Jab,fig:sameLELdiffCat,fig:frustration_explain}, where the system size is smaller to ensure better visualization). 
We explore the configurations of the system by changing the particles' orientations and positions. We index the equilibration steps by an integer $t$. At each $t$, the configuration of the system is described by the positions and the orientations of all the particles.
We only change the configuration of at most one particle per step $t$. The energy of the system reads
\begin{equation}
    E(t) = \sum_{a\leq b}N_{ab}(t)J_{ab}
    \label{eq:hamiltonian}
\end{equation}
At each step, we perform an elementary Monte-Carlo move. We thus draw with a uniform probability which particle, or group of particles, will change configuration. With probability $1-\tau$, only a single particle changes position or orientation, and with probability $\tau$, we perform a cluster move, thus updating the position or orientation of several particles at the same time. Collective moves of particles have indeed been shown to accelerate Monte-Carlo sampling \cite{frenkel2002understanding, jacobs2015rational}.

\begin{figure}
    \centering
    \includegraphics[width=\linewidth]{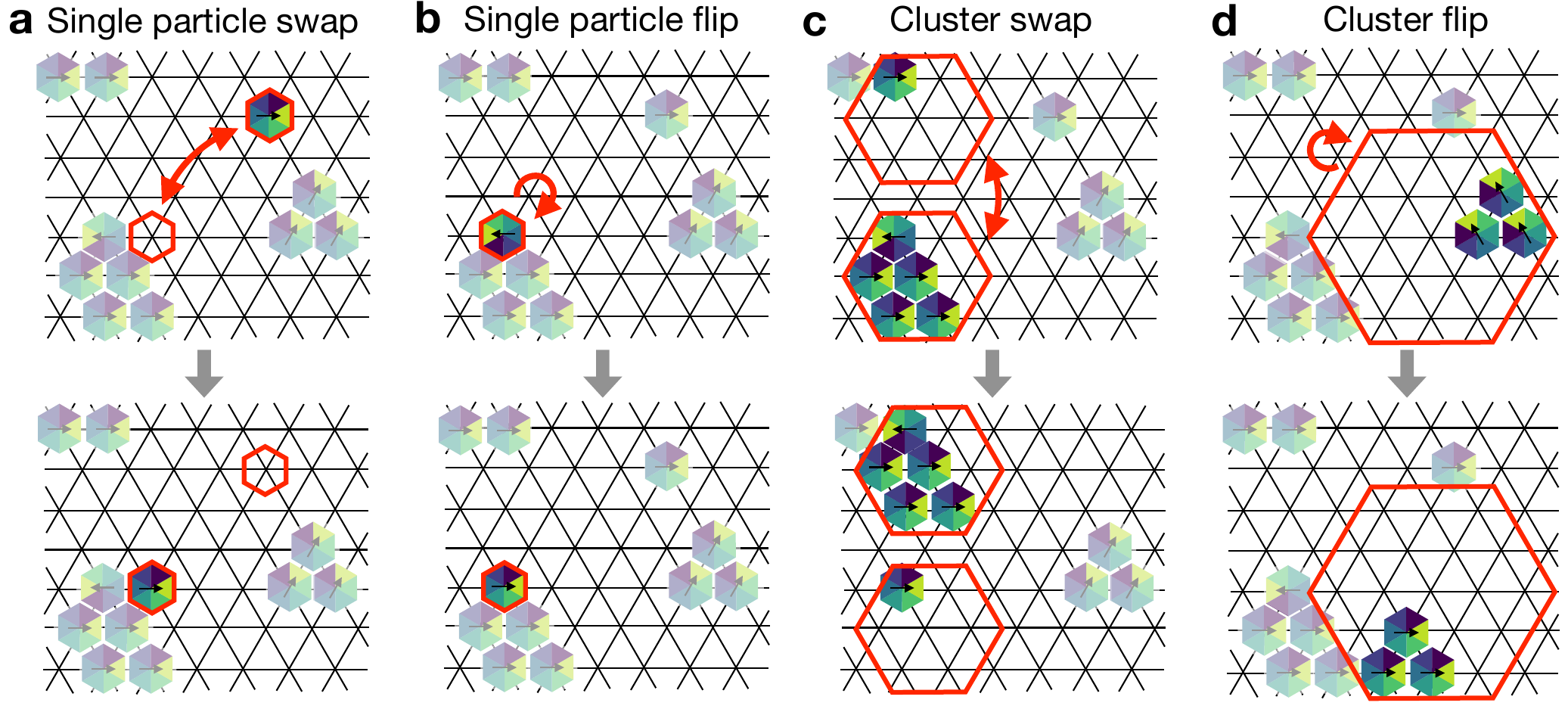}
    \caption{\textbf{We equilibrate the system by proposing moves of swap or flip of single particles or clusters of particles.} (a) A randomly chosen single particle and an empty site are swapped. (b) A single randomly chosen particle is flipped (rotated). (c) Two randomly chosen hexagonal clusters of radius $1$ are swapped. (d) One randomly chosen hexagonal cluster of radius $2$ is flipped. }
    \label{fig:cluster_move}
\end{figure}

With probability $(1-\tau)\times 1/2$, the chosen particle changes orientation. The new orientation is drawn uniformly among the orientations that are different from the current one. With probability $(1-\tau)\times 1/2$, the particle changes position on the lattice. The new position is drawn uniformly among the empty sites of the lattice. Therefore, the particles do not only diffuse to neighboring sites but rather teleport to arbitrarily distant available sites, which favors faster equilibration. We illustrate those Monte-Carlo moves on \cref{fig:cluster_move}a and b. 

With probability $\tau \times 1/2$, we attempt a swap of two site clusters. We first randomly draw two sites of the lattice, and the radius $r$ of the hexagonal shaped cluster of sites. We choose a cluster of radius $r$ with a probability scaling as $1/r^2$. In the example \cref{fig:cluster_move}(c), the chosen clusters are of radius $r=1$. The radius of the cluster is between $r=1$ ($7$ lattice sites) and $r=7$ ($169$ lattice sites). The two clusters have the same size. If they overlap, we do not attempt the move. Otherwise, two groups of sites are swapped, and the orientations of the particles on the non-empty sites are conserved. With probability  $\tau \times 1/2$, we attempt a rotation of a cluster of sites, as in \cref{fig:cluster_move}(d). The size of the cluster is chosen randomly, as well as its new orientation.  

At each step, we compute the new energy of the system $E'$ and compare it to the old energy $E$. In practice, we only recompute the energy from the bonds of the moved particle(s) and its old and new neighbors. The move is accepted according to a Metropolis criterion with temperature $T$ (always accepted if $E'<E$, accepted with a probability $p=\mathrm{exp}[-(E'-E)/(k_BT)]$ if $E'>E$).

We minimize the free energy of each system using simulated annealing. We perform 100 equally spaced inverse temperature steps starting at $1/(k_B T)=0$ and ending at $1/(k_BT)=1$. Within each temperature step we perform a number of Monte-Carlo steps equal to $7200\times N_\mathrm{particles}$, where $N_\mathrm{particles}$ denotes number of particles in the system. After the annealing, we perform $7200\times N_\mathrm{particles}$ Monte-Carlo steps at $k_BT=1$ while averaging $N_{ab}$. The results presented here are averages over five repeats of the whole annealing procedure for each interaction map. We consider $9000$ distinct interaction maps, and therefore run $45000$ simulated annealing. Each of them represents $~3$ CPU-hours, and the total amount of CPU time used for this study is therefore of the order $~135 000$ CPU-hours.

\subsection{Machine learning classification}
\label{sup:machine_learning}

To train the algorithm that recognizes the aggregation category, we first manually label $738$ images of equilibration results such as the ones shown in~\cref{fig:random_examples}. We train a dense neural-network to classify $80\%$ of the labeled data (the training set) and we test its performance on the rest of the labeled data (the test set). We then use this neural-network to classify the rest of the dataset that was not labeled manually. The result of the classification on all the data is shown in~\cref{fig:phase_diagram}. Here, we explain how the neural-network is built, trained, and show that it reliably classifies our data. 

Let us consider an individual interaction map, which we index by $i$ in the following. We first run an equilibration simulation for map $i$, and create an \textit{input vector} $\mathbf{X_i}$ to describe the output according to the procedure of described in \cref{subsec:input_vector}. The \emph{output vector} of the classification algorithm is an eight-component vector $\mathbf{Y}_i$ whose individual components represent the probability of the aggregate belonging to each of our eight categories. For hand-labeled aggregates, the true value $\mathbf{Y}^\textrm{true}_i$ of this vector is known. For instance $\mathbf{Y}^\textrm{true}_i=(0,1,0,0,0,0,0,0)$ for the aggregate of \cref{fig:introduce_Jab}(c)(2) because the aggregate belongs to the second aggregation category (namely crystals). The training of the algorithm consists in minimizing the distance between $\mathbf{Y_i^{\mathrm{true}}}$ and the predicted $\mathbf{Y_i^{\mathrm{pred}}}$ as described in \cref{subsec:learning}.

\subsubsection{Input vector}
\label{subsec:input_vector}

The input vector is composed of the interaction map ($21$ numbers), the density map, \textit{i.e.} the proportion of each types of interaction, included the empty-empty and empty full interactions ($28$ numbers), and of geometric indicators: the averaged size, the averaged number of vacancies per particles in an aggregate, and the averaged surface to volume ratio of each aggregate ($3$ numbers). The density map, the geometric indicators and the orientational orders are averaged over $5$ different simulated annealing of the system. These numbers are usually referred to as \textit{features}. 

For each hand-labeled interaction map, we multiply input vectors with the same label to take into account the symmetries of the system. Indeed, the interaction map and density map are defined up to the relabeling of the angles $\phi_i$ of~\cref{fig:introduce_Jab}(a). This relabeling corresponds to a cyclic permutation of the lines and the columns in the interaction map. 
Two interaction maps $J$ and $J'$ are thus physically equivalent if they verify
\begin{equation}
    J' = P^k \cdot J \cdot P^{-k} \text{ with } P = \begin{pmatrix} 
    0&1&0&0&0&0\\
    0&0&1&0&0&0\\
    0&0&0&1&0&0\\
    0&0&0&0&1&0\\
    0&0&0&0&0&1\\
    1&0&0&0&0&0
    \end{pmatrix} \text{ and } k\in \llbracket0,6\rrbracket
\end{equation}
Similarly, two interaction matrices $J$ and $J'$ are equivalent up to a mirror transformation of the particle if they verify 
\begin{equation}
    J' =  M \cdot J \cdot M^{-1} \text{ with } M = \begin{pmatrix} 
    0&0&0&0&0&1\\
    0&0&0&0&1&0\\
    0&0&0&1&0&0\\
    0&0&1&0&0&0\\
    0&1&0&0&0&0\\
    1&0&0&0&0&0
    \end{pmatrix}
\end{equation}

For each interaction map, we enumerate the cyclic permutation and mirror transformation of the interaction map, and add the $11$ corresponding distinct input vectors to the dataset. As a consequence, the neural-network learns that two systems are equivalent after this transformation, and the classification does not depend on the arbitrary choice of the permutation of the interaction map. It also has the advantage of increasing the number of data by a factor $12$, without having to run more simulations or classify more images.

We label $93$ systems as monomers, $53$ as oligomers,  $59$ as fibers, $50$ as gels, $159$ as sponges, $159$ as crystals, $114$ as polycrystals and $51$ as liquids. 
Those examples are spread among all values for affinity and asymmetry ($\mu$ and $\sigma$). For this dataset, we have the input vector $\mathbf{X}$, and the true label vector $\mathbf{Y}^{\mathrm{true}}$ which collects the values of all $\mathbf{Y}_i^{\mathrm{true}}$.
The final total input of the algorithm $\mathbf{X}$ is composed of $738 \times 12$ input vectors, each containing $21+28+3=52$ features. We normalize each feature by its average value over the whole dataset.

\subsubsection{Neural-network}
\label{subsec:learning}

Here, we choose the structure of the neural-network and the learning parameters such that the accuracy of the prediction on both the training and the test set are close to $100\%$. As is often the case in machine learning, these choices are arbitrary and another neural-network architecture could give similar results~\cite{mehta2019high}. 

To transform the input $\mathbf{X}$ into the predicted labels $\mathbf{Y}^{\mathrm{pred}}$, we use a dense network of $5$ layers implemented with the python library \textit{keras}. Each layer is respectively composed of $100$, $200$, $400$, $100$ and $30$ neurons. At each layer, we use the rectified linear unit function.
The network is trained by minimizing the cross entropy loss function. The optimizer is \textit{adam}. We also implement an L1 and L2 regularization, with factors $10^{-4}$ and $10^{-5}$. We optimize the loss function for $1500$ iteration (epochs) on different batches of $128$ data (minibatches). We measure a training accuracy of $99.7\%$ and a test accuracy of $99.3\%$ on the labeled dataset. 

With this neural-network, we then classify the rest of the dataset that was not labeled manually. We ensure that the labeled and unlabeled data have comparable distribution by labelling a sufficient amount of data for all values of the affinity and asymmetry, and in each aggregation category. We thus label between $7$ and $26$ aggregates out of the $200$ data for each value of the couple ($\mu$, $\sigma$).

\subsection{Frustration and naive minimization}
\label{sup:gas_faces}

As a baseline for the physics of aggregation in the absence of geometrical constraints, we determine the proportion of each face pair in a system where their paring is completely unconstrained, save for the conservation of the number of bonds and number of particles. The free-energy of such a system reads 
\begin{equation}
F(\{N_{ab}\}) = \sum_{a \leq b} N_{ab}J_{ab} -k_BT \sum_{a \leq b} \frac{N_{ab}}{N_\mathrm{bonds}} \Bigl(\mathrm{ln} \frac{N_{ab}}{N_\mathrm{bonds}} - 1\Bigl),
\end{equation}
where the indices $a$ and $b$ run between 0 and 6 and the faces labeled ``0'' refer to an empty site in the following.
We ensure the conservation of the number of bonds and the number of faces are ensured with $7$ Lagrange multiplier ($\lambda$ and $\{\lambda_a\}_{a\in \llbracket 1,6 \rrbracket} $). We therefore solve the following set of equations numerically for $k_BT=1$.
\begin{align*}
    \begin{cases}
    &N_{ab}^{\mathrm{(min)}}= \frac{\partial}{\partial N_{ab}} \biggr[ F(\{N_{ab}\}) + \lambda \Bigl( \sum_{ a\leq b} N_{ab}- N_\mathrm{bonds} \Bigl) + \sum_{a\geq 1} \lambda_a \Bigl(\sum_{b}N_{ab}+N_{aa} - N_\mathrm{particles} \Bigl) \biggr]\\ \numberthis 
    \label{eq:naive_minimization}   \\
    &\sum_{\leq a \leq b} N_{ab} = N_\mathrm{bonds} \\ 
    &\sum_{b}N_{ab}+N_{aa} = N_\mathrm{particles},
    \end{cases}
\end{align*}
We then measure the frustration as the positive energy difference between the minimal energy resulting determined from the Boltzmann distribution, and the equilibrium energy determined in the numerical simulation. 
\begin{equation}
\label{eq:frustration}
    \Delta E_{f}=\sum_{a \leq b} (N^{\mathrm{(s)}}_{ab}- N^{\mathrm{(m)}}_{ab} )J_{ab}
\end{equation}
This energy difference is due to the favored interactions that could not be realized in the numerical simulation, because they lead to an extra interaction that is not accounted for in the energy minimization. We then compute the relative energy difference, by dividing frustration by $E$, the energy of the system measured in the simulation.
\begin{equation}
\label{eq:relative_frustration_supp}
    f=\Delta E_{f} /E
\end{equation}
We compare the frustration in dense and dilute conditions. We first measure the frustration $f_\mathrm{dilute}$ (\cref{eq:relative_frustration_supp}) for the 9000 simulated annealing results introduced in the previous section. These simulations are in dilute conditions ($500$ particles in a lattice of $60\times60$ sites). We then perform simulated annealing for the same 9000 interaction maps in dense conditions: $484$ particles in a lattice of $22\times22=484$ sites, while conserving the annealing protocol: $100$ temperature steps and $7200$ Monte-Carlo steps per temperature and per particle in the system. Finally, we compute the associated frustration $f_\mathrm{dense}$ as before. Because the measure of frustration is normalized by the number of particles, the small difference in the total number of particles in these calculations is irrelevant. We then compare $f_\mathrm{dense}$ and $f_\mathrm{dilute}$ for the 9000 interaction map (\cref{fig:stat_frustration}). We conclude that forming sparse or dimensionally reduced systems is a way to reduce frustration when it is high in the dense conditions.

\subsection{Evaluating a measure of the interaction with machine learning}
\label{subse:method_interpretation}

Here, describe our use of machine learning to predict the outcome of aggregation from partial information on an interaction map, as in \cref{fig:interpretationML}. As described in \cref{subsec:learning}, we have at our disposal a list of labeled interaction maps, each assigned to an aggregate category.
We train a neural-network to predict the aggregate category from partial information on the interaction map following the procedure described in \cref{subsec:input_vector}, with the exception that we use only half of our sample of 9000 system for computational speed.
We denote the vector containing the partial information by $\mathbf{X}^{(\text{partial})}$. We then assess its accuracy, as reported on the vertical axis of \cref{fig:interpretationML}.
To ensure that our result does no depend on our choice of training sample, we redistribute the data into the training and test sets and repeat this process $20$ times and report the average result. The specific subset of 4500 systems out of 9000 is redrawn randomly each time, but always includes the $738$ manually labeled data, the other being labeled with the method described in \cref{sup:machine_learning}.

For each initial dataset $\mathbf{X}^{(\text{partial})}$, we keep the architecture of the neural-network unchanged ($6$ layers composed of $100$, $200$, $500$, $200$, $100$ and $30$ neurons respectively) and use the same training procedure as in ~\cref{sec:classification}. The rest of the learning parameters, such as the regularization factors or the number of epochs, are identical to that of the network described in~\cref{sup:machine_learning}.

\subsection{Measure of the propagability}
\label{subsec:propagability}

Here, we describe our procedure to compute the propagability from the interaction map, by enumerating the possible periodic lines of the particles and measuring their energy per particle. 
For a given initial orientation $\varphi_0$, and a given periodicity $n$, we enumerate all $6^{n-1}$ orientations $\{ \varphi_k \}$ of the particles such that a line is in the configuration ($\varphi_0,\varphi_1, .. \varphi_{n-1}, \varphi_0$), which we refer to as a periodic motif.

The effective coupling for a given initial orientation $\varphi_0$ and a given periodicity $n$, which we denote by $J^{\mathrm{eff}}(n, \varphi_0)$ is the minimal possible energy for a periodic motif over our enumeration of its orientations.
\begin{equation}
\label{eq:jeff_line}
J^{\mathrm{eff}}(n,\varphi_0) = \underset{\varphi_1, ... \varphi_{n-1}}{\mathrm{min}}\frac{J_{\varphi_0 \varphi_1}+ J_{\varphi_1 \varphi_2}+... + J_{\varphi_{n-1} \varphi_0}}{n-1} 
\end{equation}
From a given interaction map, the computation of the values of $J^{\mathrm{eff}}(n,\varphi_0)$ is a straightforward operation on the entries of the matrix. Because of the rotation invariance of the system, we only compute this value for $\varphi_0=0$, $\pi/3$ and $2\pi/3$. We also only compute this number for $n\leq 6$, because since the particles only has six orientations, there cannot be any most favorable one-dimensional periodic motif of more than $6$ particles. Computing one value for $J^{\mathrm{eff}}$ is at maximum an enumeration of $6^5$ configurations, which is accessible numerically.

For a given periodicity $n$, the organization of the gray particles of  \cref{fig:propag_example}(a) such that each of the energies $e_1(n)$,$e_2(n)$ and $e_3(n)$ are minimum is then simply computed from the effective interactions of equation~\ref{eq:jeff_line}:
\begin{equation}
\mathbf{J}^{\mathrm{eff}}(n) = \left(J^{\mathrm{eff}}(\varphi_0=0, n),J^{\mathrm{eff}}(\varphi_0=\pi/3,n), J^{\mathrm{eff}}(\varphi_0=2\pi/3,n) \right)
\end{equation}
Here, we do not count the interactions with the particles that may or may not be present between the gray lines of \cref{fig:propag_example}(b). The number of such particles for the largest motifs considered here is indeed well beyond our ability to enumerate them. 

We then choose the best periodicity $n^*$ to be the one where the minimum of the three line energies is the lowest $n^* =\underset{n, i}{\textrm{argmin}} [e_i(n)]$ An alternative definition of the propagability where
$n^*=\underset{n}{\textrm{argmin}}[e_1(n)+e_2(n)+e_3(n)]$
leads to a poorer prediction accuracy.
The propagability is thus defined as the list of six features, all computed from the interaction map $J$: the three components of the optimal effective coupling vector $\mathbf{J}^{\mathrm{eff}}(n^*)=[e_1(n^*),e_2(n^*),e_3(n^*)]$, the optimal periodicity $n^*$, the particle affinity $\text{av}(J)$ and asymmetry $\text{std}(J)$. Because we evaluate together periodic lines of identical length $n$, we authorize some of those lines to be of period $n/2$ or $n/3$, to match a longer and favored periodic line. For instance, in \cref{fig:propag_example}e, the second motif of the polycrystal is of period $1$, but the motif of lowest energy is the first line, of period $2$. For this reason, we could consider motifs of length larger than $6$, to allow for commensurate motifs of period $4$ and $5$ for instance. Yet, we reason that we did not observe aggregate periodicity larger than $6$, and that this computation is not accessible numerically.

\section{Supplementary material}
\label{sec:SI}

\subsection{Choice of an interaction map with vanishing surface energy}
\label{sup:invariance}

In the interaction map introduced in~\cref{fig:introduce_Jab}, we only account for the interaction energy between the faces of two particles. In principle, we could also define an interaction energy between a particle face and an empty site of the lattice, or between two empty sites of the lattice. Here, we show that in a system with a fixed number of particles these interactions can be set to any arbitrary value without loss of generality.

We refer to sites of the lattice where a particle is present as \emph{full}, and those where there is no particle as \emph{empty}. While we label \textit{full-full interaction} using the notation $J_{ab}$ as in the main text, we denote the energy of an \textit{empty-full interaction} by $J_{a0}$ and that of an \textit{empty-empty} interaction by $J_{00}$. 

In the general case of non-zero $J_{a0}$ and $J_{00}$, we generalize~\cref{eq:hamiltonian} to write the total energy of the system as
\begin{equation}
    E = J_{00}N_{00}+\sum_{a=1}^{N_{\text{faces}}} J_{a0}N_{a0} + \sum_{a=1}^{N_{\text{faces}}}\sum_{b=1}^{a}J_{ab}N_{ab}
\end{equation}
Hexagonal particles give rise to $N_\text{faces}+1=7$ conserved quantities. These quantities are the total number of bonds, and the number of $a$-faces:
\begin{align}
\label{eq:Nbonds_cons}
N_{\mathrm{bonds}} &= N_{00}+\sum_{a=1}^{N_{\mathrm{faces}}} N_{a0} + \sum_{a=1}^{N_{\mathrm{faces}}}\sum_{b=1}^{a}N_{ab} \\
\label{eq:Nfaces_cons}
N_a &=  N_{a0} + \sum_{b\neq a}N_{ab} + 2 N_{aa} 
\end{align}
and since each particle has one of each type of faces, we have
\begin{equation}
    \forall a\in\llbracket{1,N_\text{faces}}\rrbracket \qquad N_a=N_\text{particles}
\end{equation}
We may thus subtract a linear combination of $N_\text{bonds}$ and $N_a$ from the system energy $E$ without changing the physics of the system. We specifically choose a new shifted energy
\begin{align*}
E' &= E - J_{00} N_{\mathrm{bonds}} - \sum_{a=1}^{N_{\mathrm{faces}}}(J_{a0}-J_{00})N_{\mathrm{particles}}\\
 &= \sum_{a=1}^{N_{\mathrm{faces}}}\sum_{b=1}^{a} (J_{ab}-J_{00}  -(J_{a0}-J_{00})-(J_{b0}-J_{00}))N_{ab}\\
 &= \sum_{a=1}^{N_{\mathrm{faces}}}\sum_{b=1}^{a} (J_{ab}+J_{00} -J_{a0}-J_{b0})N_{ab}\\
 &= \sum_{a=1}^{N_{\mathrm{faces}}}\sum_{b=1}^{a} J'_{ab} N_{ab} \numberthis \label{eq:energy_shift}
\end{align*}
which boils down to the definition of a new interaction map $J'_{ab}=J_{ab}+J_{00} -J_{a0}-J_{b0}$ for which the energies of the full-empty and empty-empty bonds is zero, as assumed throughout the main text.

\subsection{Equilibration}
\label{sup:equilibration}

To demonstrate that our simulations result in well-equilibrated systems, here we show that with our annealing protocol the measured final composition of a system and aggregate morphology are independent of the number of annealing steps, particle density, number of particles, and percentage of cluster move in the annealing, if we go beyond the chosen simulation parameters. 

\begin{figure}
    \centering
    \includegraphics[width=\linewidth]{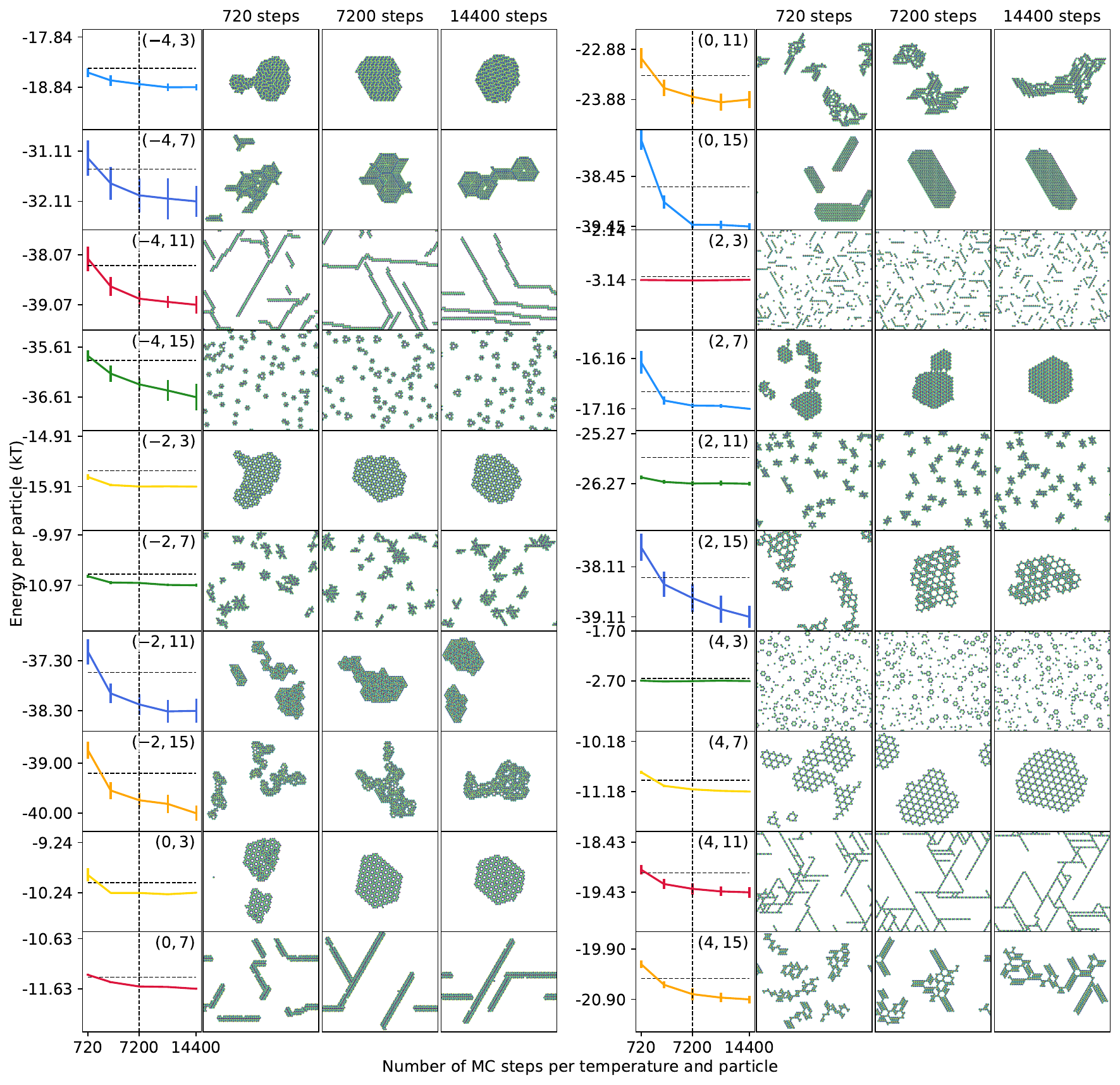}
    \caption{\textbf{Our final systems are well equilibrated.} Beyond 7200 steps per temperature and lattice sites (horizontal dashed line), the energy and the system configuration are unchanged by increasing the annealing time. Equilibrating curve and system snapshot at different number of steps. The horizontal dashed line is a relative energy difference of $2\%$ with the energy measured for the maximal number of steps. The vertical scale is $2 kT$ for all plots.  For each system, the interaction maps are similar to that of~\cref{fig:random_examples}, and the values of affinity and asymmetry $(\mu, \sigma)$ are indicated in the top right of each energy profile. Error bars indicate the standard error averaged over $10$ simulations.} 
    \label{fig:equilibration}
\end{figure}

For the interaction maps shown in~\cref{fig:random_examples}, we measure the energy per particle as a function of the number of Monte-Carlo steps performed per temperature and per lattice site. The results are shown in~\cref{fig:equilibration}, together with images of an equilibrium configuration at different time steps. As expected, the energy per particle decreases with the duration of the equilibration, up to a limit after which increasing the number of steps does not decrease the energy. We choose the number of steps for the simulation to be such that the relative lowering of the energy resulting from a doubling of the number of steps is smaller than $2\%$. We find that this result can be obtained by performing $14400$ Monte-Carlo steps per temperature and per particles. This threshold corresponds to the horizontal dashed line on the energy evolution on~\cref{fig:equilibration}. The images on the figure confirm that the configuration of the system also does not change by increasing the number of steps to $14400$ per temperature step per site. 

\begin{figure}
    \centering
    \includegraphics[width=\linewidth]{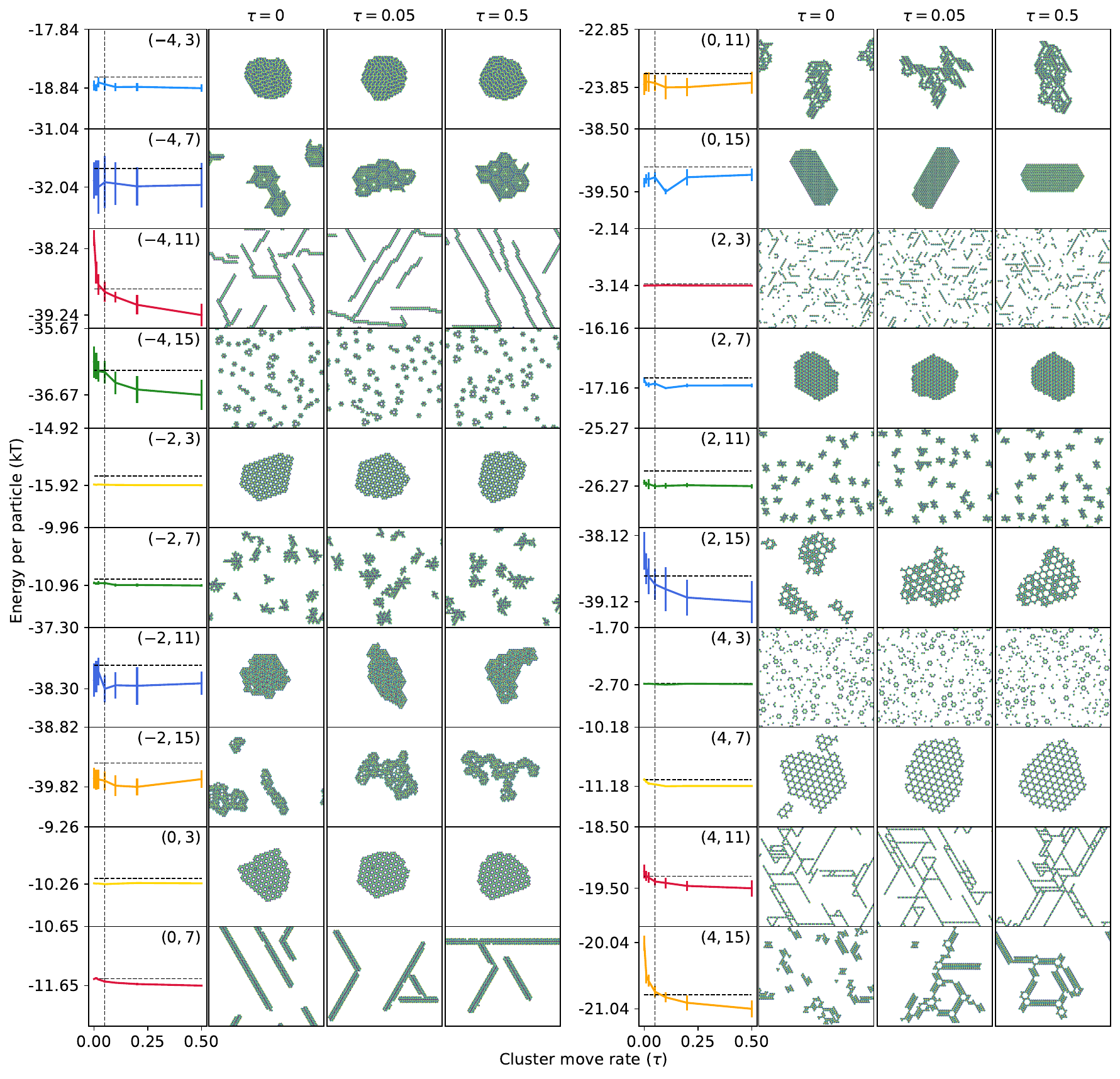}
    \caption{\textbf{The aggregate morphology is independent of the rate of cluster moves during the annealing} Beyond $5\%$ of cluster moves during the annealing (vertical dashed line), the energy, and the system configuration are unchanged by increasing the rate of cluster moves. Equilibrating curve and system snapshot at different number of cluster move rate. The horizontal dashed line is a relative energy difference of $5\%$ with the minimum energy measured. The vertical scale is $1.5 kT$ for all plots. 
    For each system, the interaction maps are similar to that of~\cref{fig:random_examples}, and the values of affinity and asymmetry $(\mu, \sigma)$ are indicated in the top right of each energy profile. Error bars indicate the standard error averaged over $10$ simulations.}
    \label{fig:rate_dependence}
\end{figure}

We then justify the need to implement a small percentage of cluster moves in the simulated annealing to correctly determine the system's equilibrium configuration. We compare the energy per particle and aggregate morphology for systems that were equilibrated with a rate $\tau$ of cluster moves (as explained in \cref{fig:cluster_move}). We show these results in \cref{fig:rate_dependence}. For a few systems, such as $(-4,11), (2,15), (4,15)$, the energy per particle is smaller of a few tens of $k_BT$ when the rate of cluster move is non-zero. Moreover, aggregates cluster moves allow aggregates of small sizes to assemble into gels, such as the systems $(0,11)$ and $(4,15)$. This happens only if there are some attractive interactions between the small aggregates, and on the contrary, implementing cluster moves does not change the equilibrium morphology of aggregate that we label as oligomers, such as $(-2,7)$ or $(2,11)$. In all our study, we therefore chose a rate of $5\%$ cluster moves during the simulated annealing, to guarantee that we are closed to the equilibrium configuration.

\begin{figure}
    \centering
    \includegraphics[width=\linewidth]{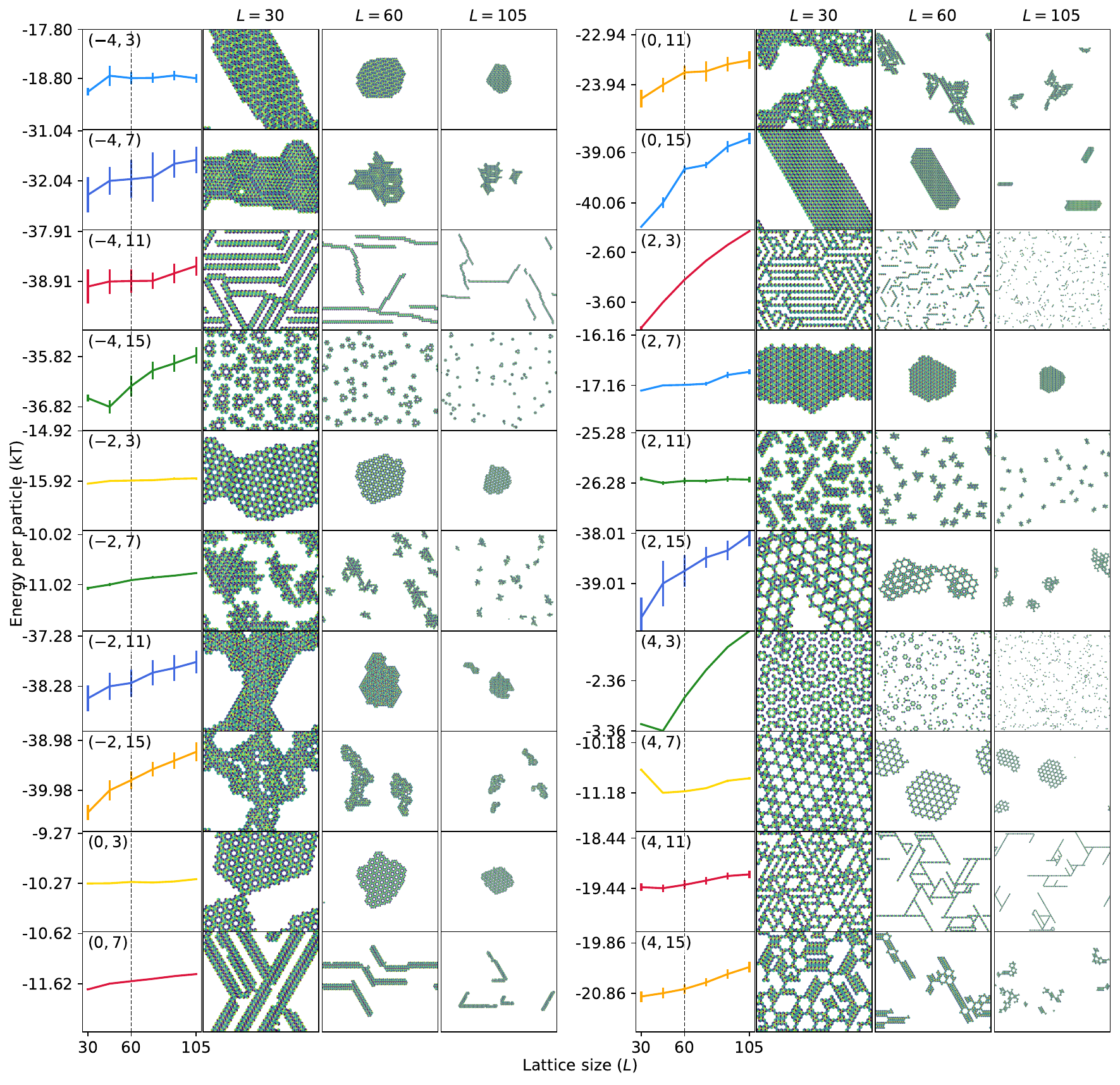}
    \caption{\textbf{In dilute systems, the aggregate morphology is independent of the particle density.} 
    We vary the size of the system ($N_\mathrm{sites}=L\times L$) while keeping the number of particles constant at $500$. For each system, the interaction maps are similar to that of~\cref{fig:random_examples}, and the values of affinity and asymmetry $(\mu, \sigma)$ are indicated in the top right of each energy profile. We show the energy per particles as a function of the lattice size. In most cases, the energy does not vary with the system size. For the system of low energy, such $(\mu, \sigma)=(2,3), (4,3), (4,7)$, the entropic contribution is sufficiently important for the energy to increase with system size. The vertical dashed lines corresponds to the chosen number system size for this study, \textit{i.e.} $60\times60$ lattice sites.
    The vertical scale is $2 kT$ for all plots. The number of Monte-Carlo steps of the annealing is always $7200\times N_\mathrm{particles}$.
    }
    \label{fig:density_dependence}
\end{figure}

We study the influence of the density of particles on the aggregate morphology in \cref{fig:density_dependence}. We vary the size of the system ($N_\mathrm{sites}=L\times L$), and keep the number of particles $N_\mathrm{particles}=500$ constant. When the system is of density one ($L=22$), the energy per particle can be very different from the dilute systems, as discussed in relation with \cref{fig:stat_frustration} of the main text. As illustrated in~\cref{fig:density_dependence} however, for smaller densities the energy per particle does not vary with the system size in most cases when $L$ increases. However, for aggregates with energy per particles of the order of the temperature (we recall that $k_BT =1$) such as examples (2,3) and (4,3) in~\cref{fig:density_dependence}, the energy per particle increases with the system size. This suggests that for very dilute systems, the entropic contribution to the free energy becomes larger than the enthalpic one. Except for these situations, the aggregate morphologies are not modified upon increasing the system density, and the particles' organization remains the same.

\begin{figure}
    \centering
    \includegraphics[width=\linewidth]{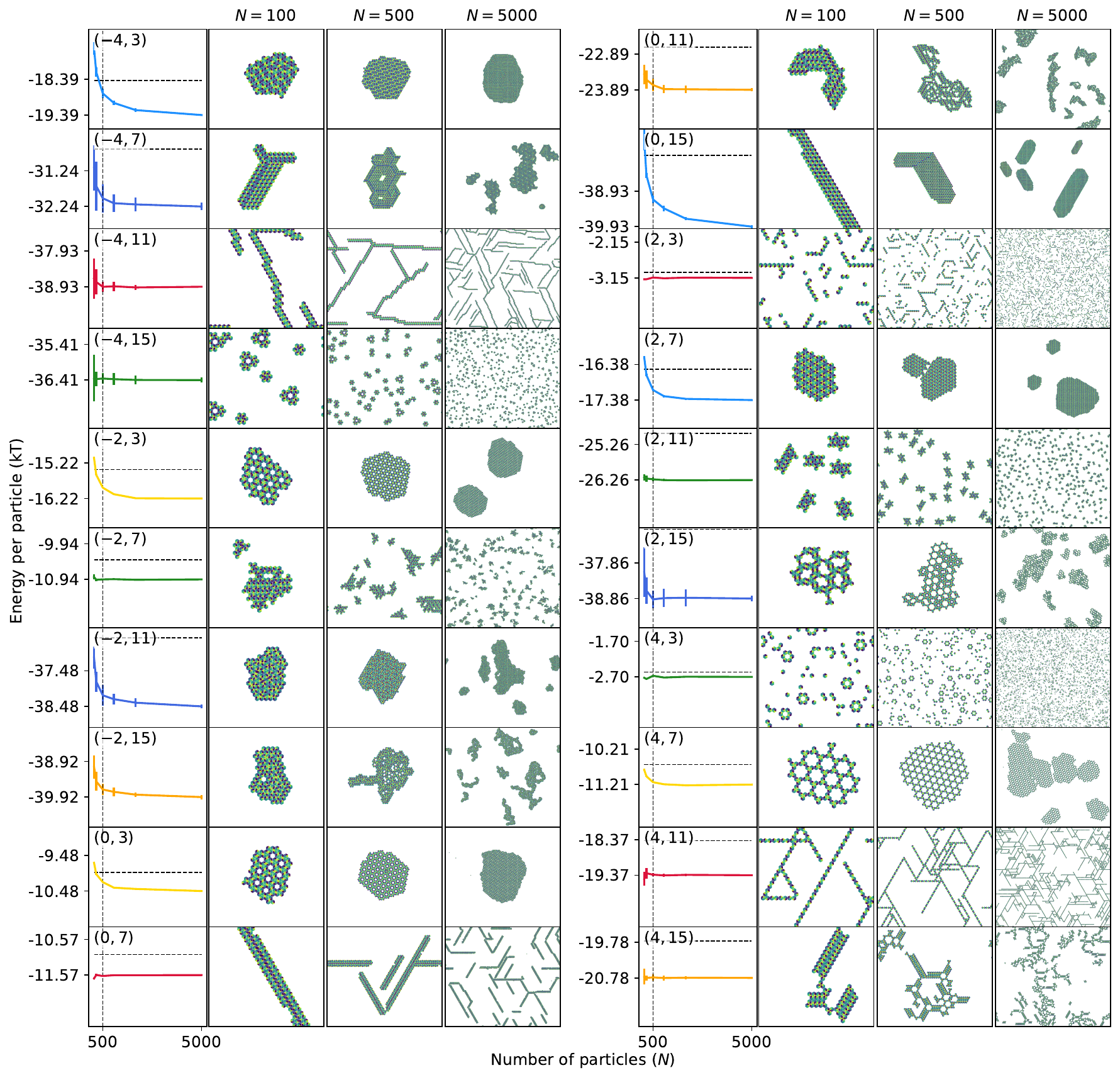}
    \caption{\textbf{The aggregate morphology is independent of the number of particles.} The energy and the system configuration are unchanged by increasing the number of particles at fixed density. Equilibrating curve and system snapshot at different number of cluster move rate. The horizontal dashed line denotes a relative energy difference of $5\%$ with the minimum energy measured. The vertical dashed lines corresponds to the chosen number of particles for this study, \textit{i.e.} 500 particles. The vertical scale is $2.8 kT$ for all plots.
    For each system, the interaction maps are similar to that of~\cref{fig:random_examples}, and the values of affinity and asymmetry $(\mu, \sigma)$ are indicated in the top left of each energy profile. Error bars indicate the standard error averaged over $10$ simulations.}
    \label{fig:Nparticles_dependence}
\end{figure}

Finally, we study the influence of the total number of particles on the observed aggregate shapes, while keeping a constant density and the similar annealing protocol. The examples of \cref{fig:Nparticles_dependence} reveal that increasing the number of particles up to $5000$ particle changes the system energy for dense aggregates of maximal sizes, because it decreases the relative number of particle at the surface of the aggregate (which scales like the square root of the number of particles). For systems of smaller sizes or dimension, however, the energy per particle is unchanged. More importantly, the aggregate's morphology remains identical if we increase the number of particle for all those examples. This proves that the equilibrium configuration of $500$ particles is sufficient to generically characterize the relation between the interaction map and the aggregate morphology. 

\subsection{Definition of the aggregate categories}
Here, we show that the eight aggregate categories we introduced in the main text satisfactorily describes all the morphologies resulting from the aggregation of particles with random interactions. We explain the criterion we use for the manual labelling of the data, and show that a neural-network accurately learns to recognize these categories. Finally, we show that the rare systems for which the classification is ambiguous correspond to aggregates where two morphologies coexist in the same system. 

\begin{figure}
    \centering
    \includegraphics[width=0.76\textwidth]{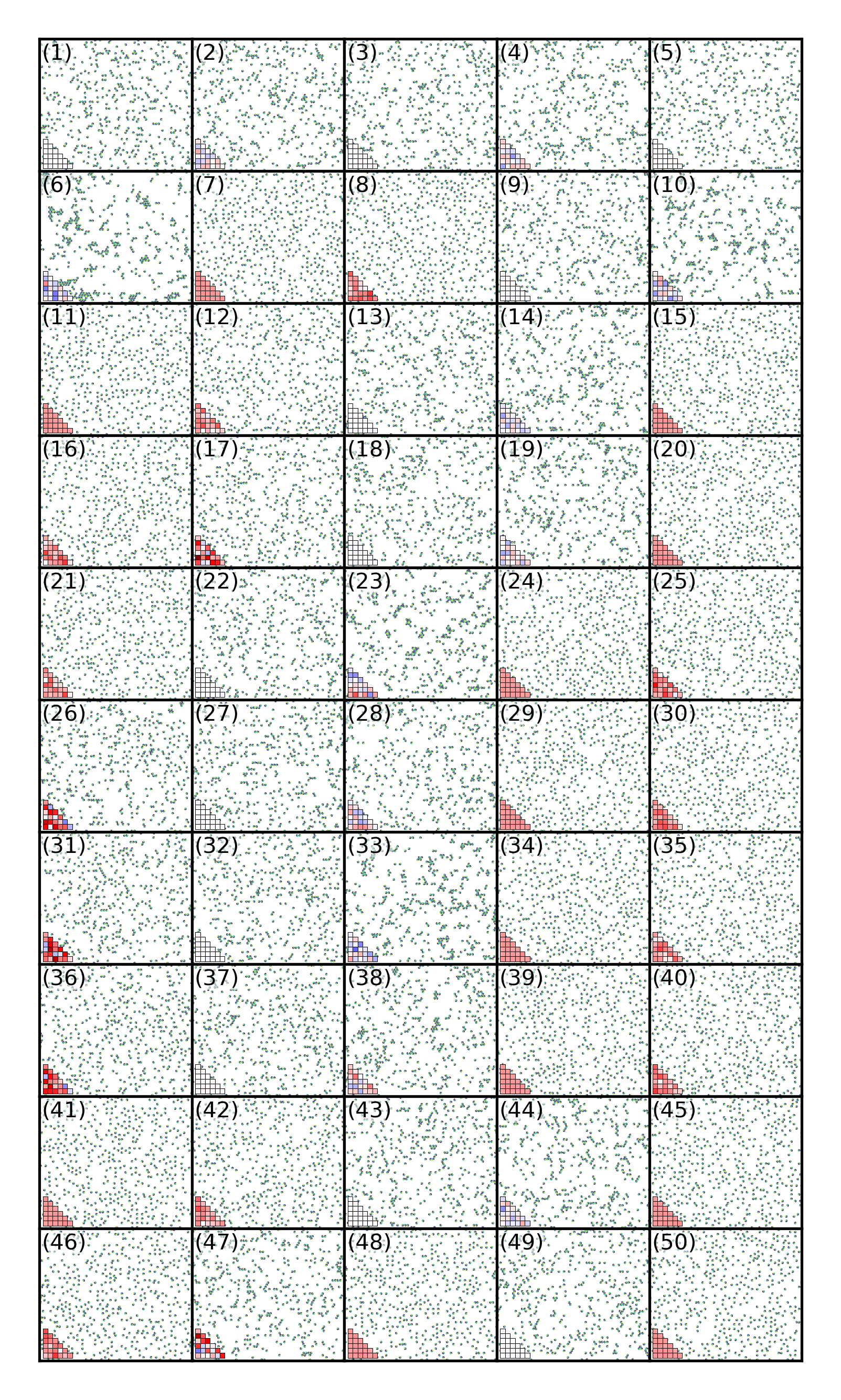}
    \caption{We label as monomers systems of non-interacting particles.}
    \label{fig:label_monomer}
\end{figure}

\begin{figure}
    \centering
    \includegraphics[width=0.76\textwidth]{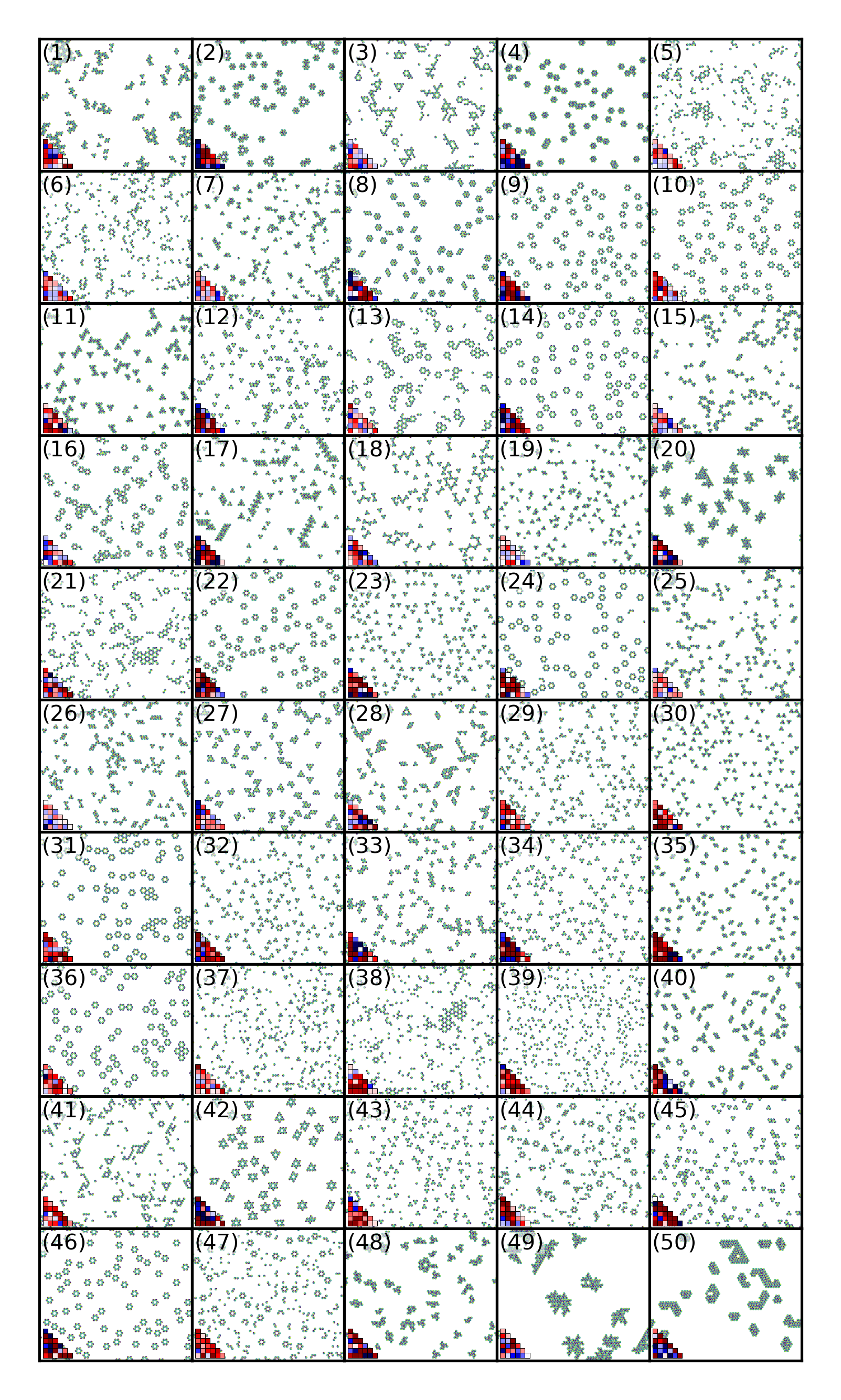}
    \caption{We label as oligomers systems of non-interacting identical aggregates.}
    \label{fig:label_oligomer}
\end{figure}

\begin{figure}
    \centering
    \includegraphics[width=0.76\textwidth]{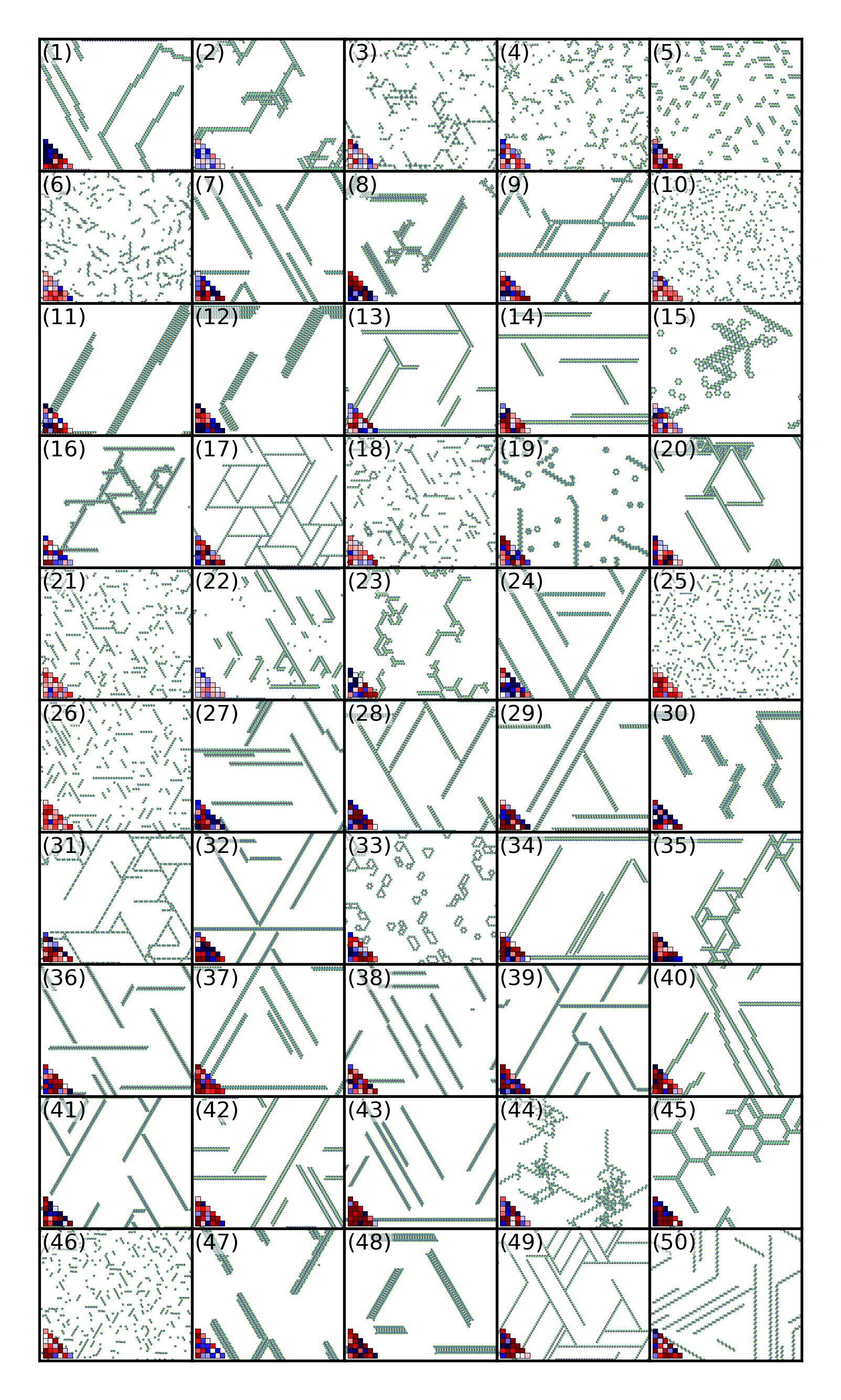}
    \caption{We label one-dimensional aggregates as fibers.}
    \label{fig:label_fiber}
\end{figure}

\begin{figure}
    \centering
    \includegraphics[width=0.76\textwidth]{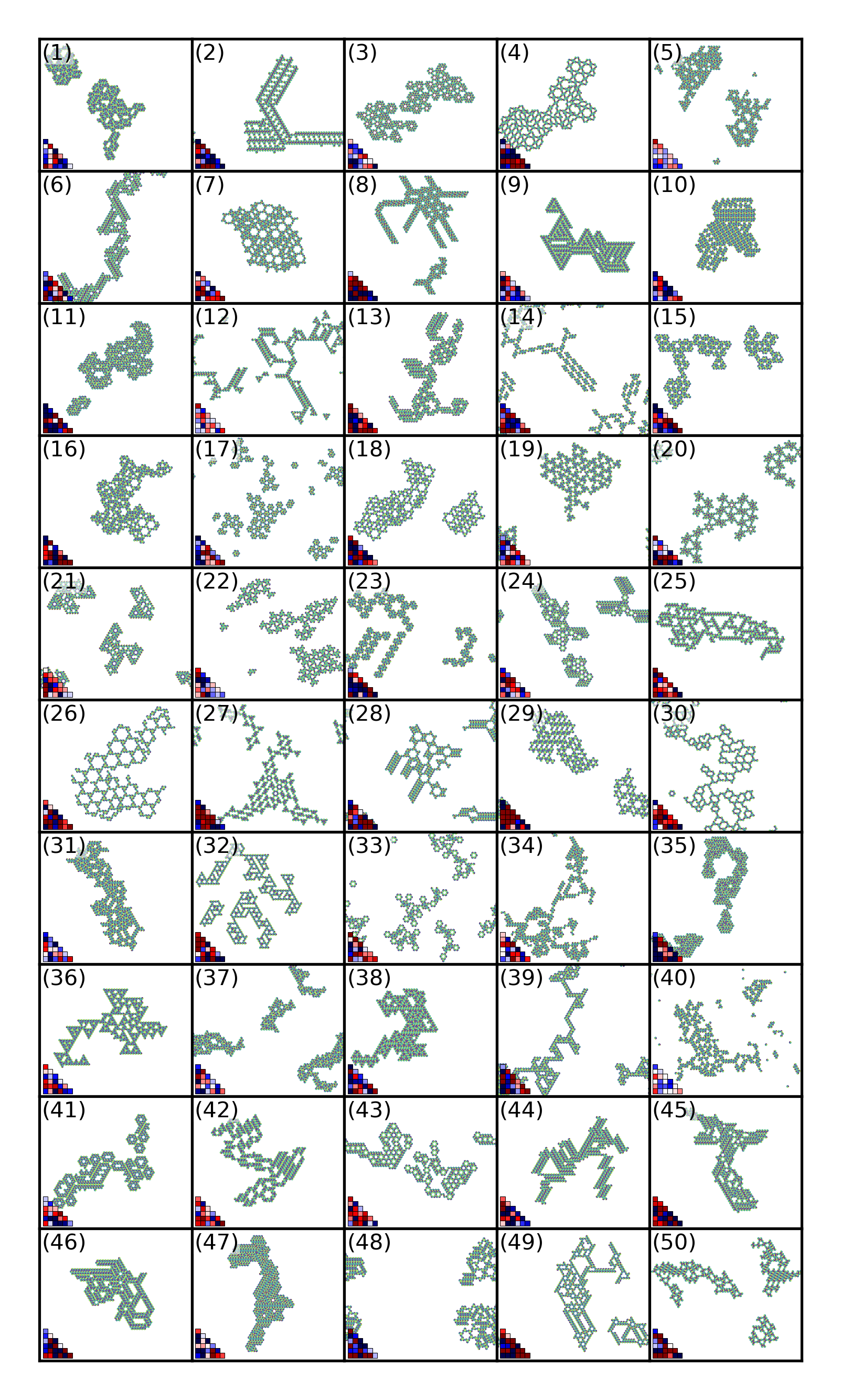}
    \caption{We label as gels aggregates of finite size that do not have orientational order across the whole aggregate, or large aggregates that have no surface tension.}
    \label{fig:label_gel}
\end{figure}

\begin{figure}
    \centering
    \includegraphics[width=0.76\textwidth]{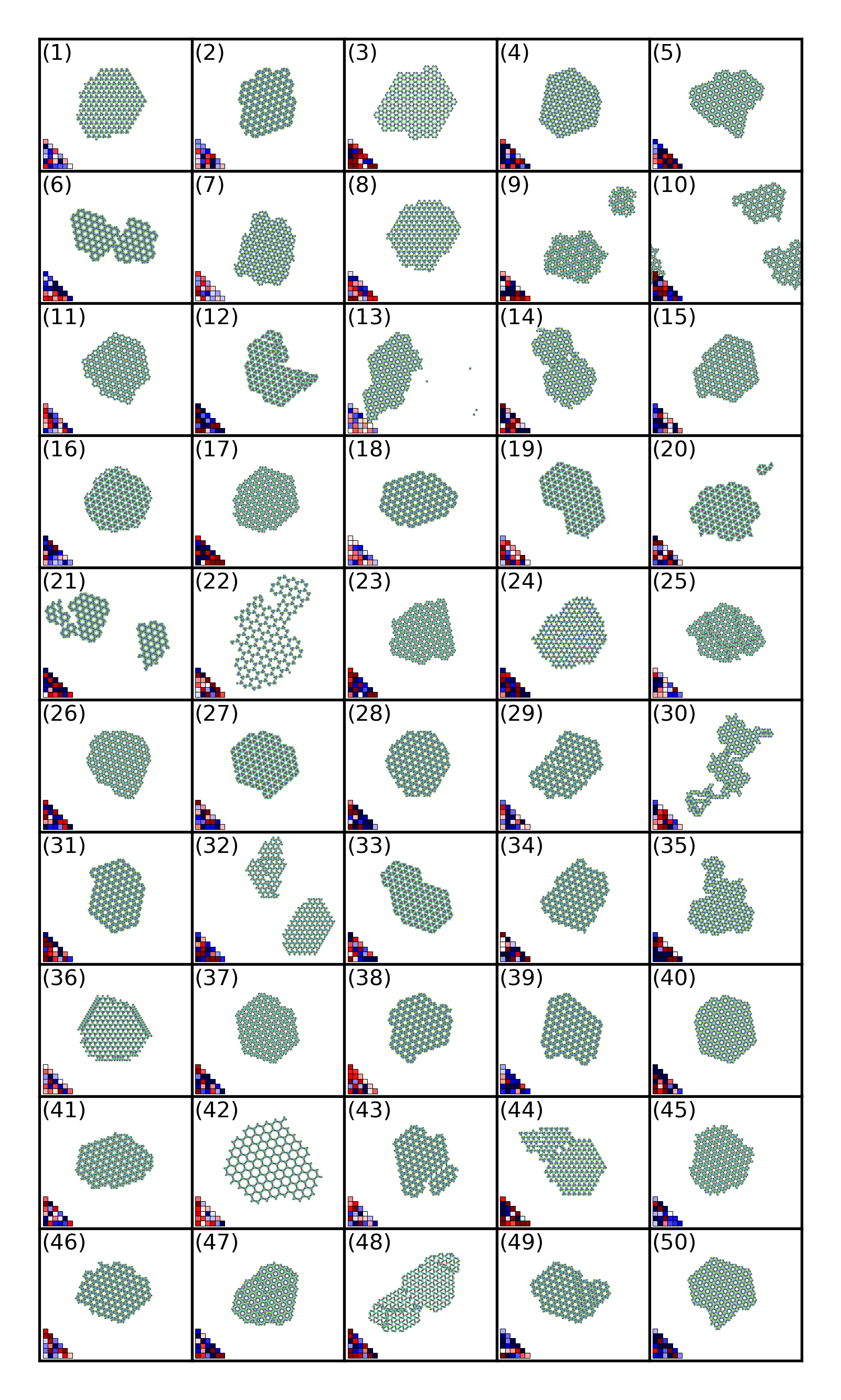}
    \caption{We label as sponges aggregates that contain all or almost all particles in the simulation, display orientational order and are porous.}
    \label{fig:label_sponge}
\end{figure}

\begin{figure}
    \centering
    \includegraphics[width=0.76\textwidth]{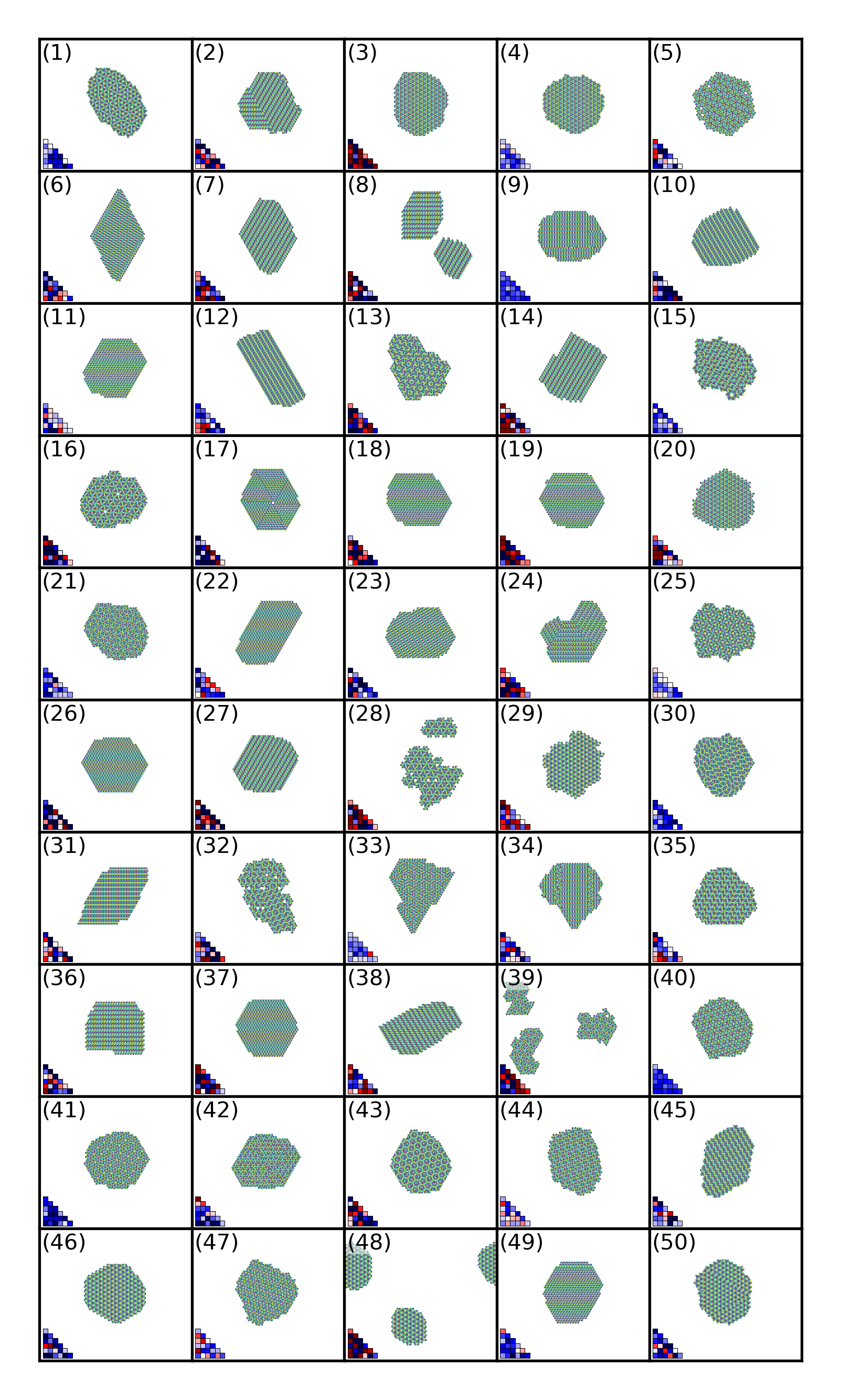}
    \caption{We label as crystals aggregates that contain all or almost all particles in the simulation, display orientational order and are not porous}.
    \label{fig:label_crystal}
\end{figure}

\begin{figure}
    \centering
    \includegraphics[width=0.76\textwidth]{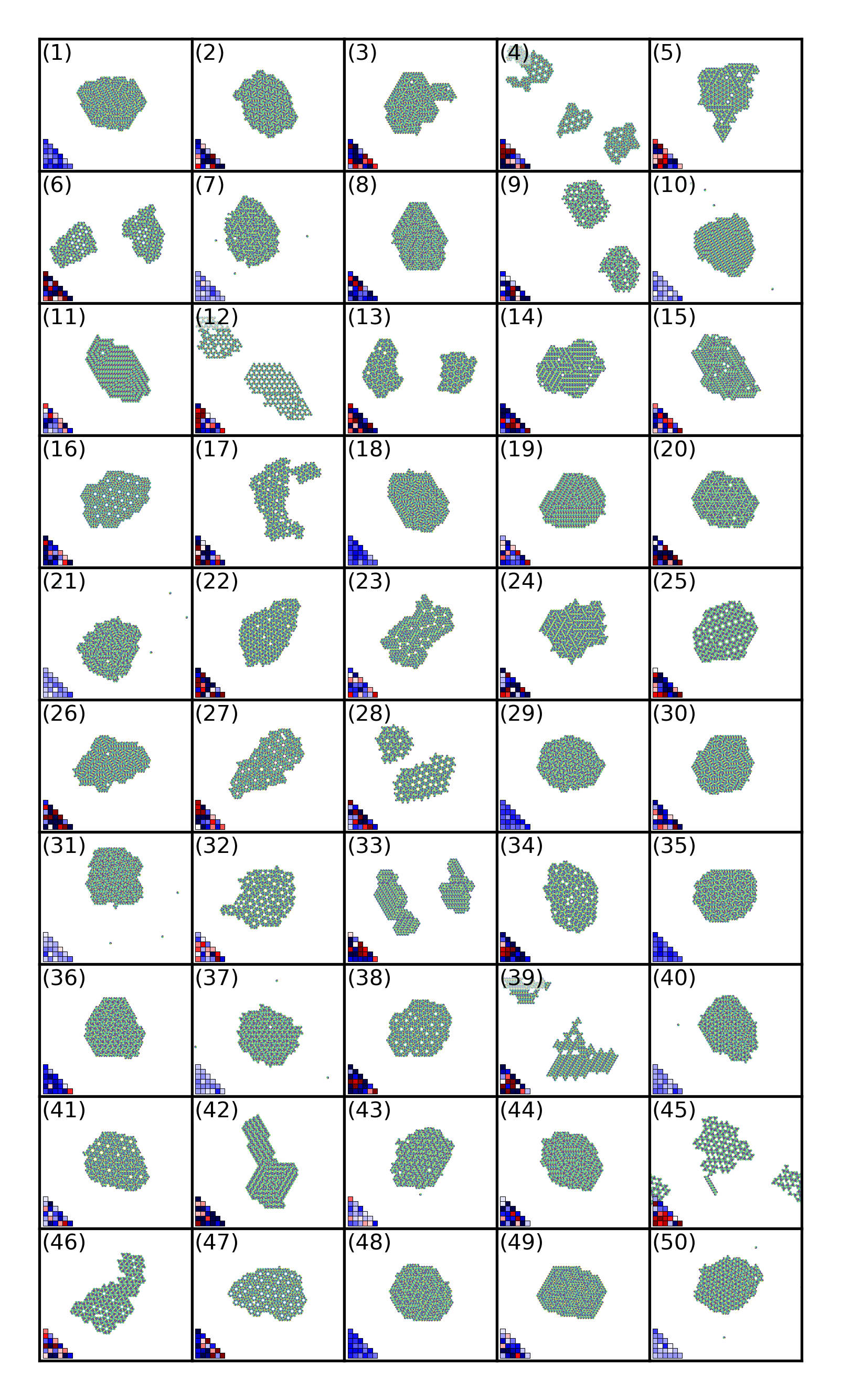}
    \caption{We label as polycrystals aggregates of short-ranged orientational order, or with several competing orientational order.}
    \label{fig:label_polycrystals}
\end{figure}

\begin{figure}
    \centering
    \includegraphics[width=0.76\textwidth]{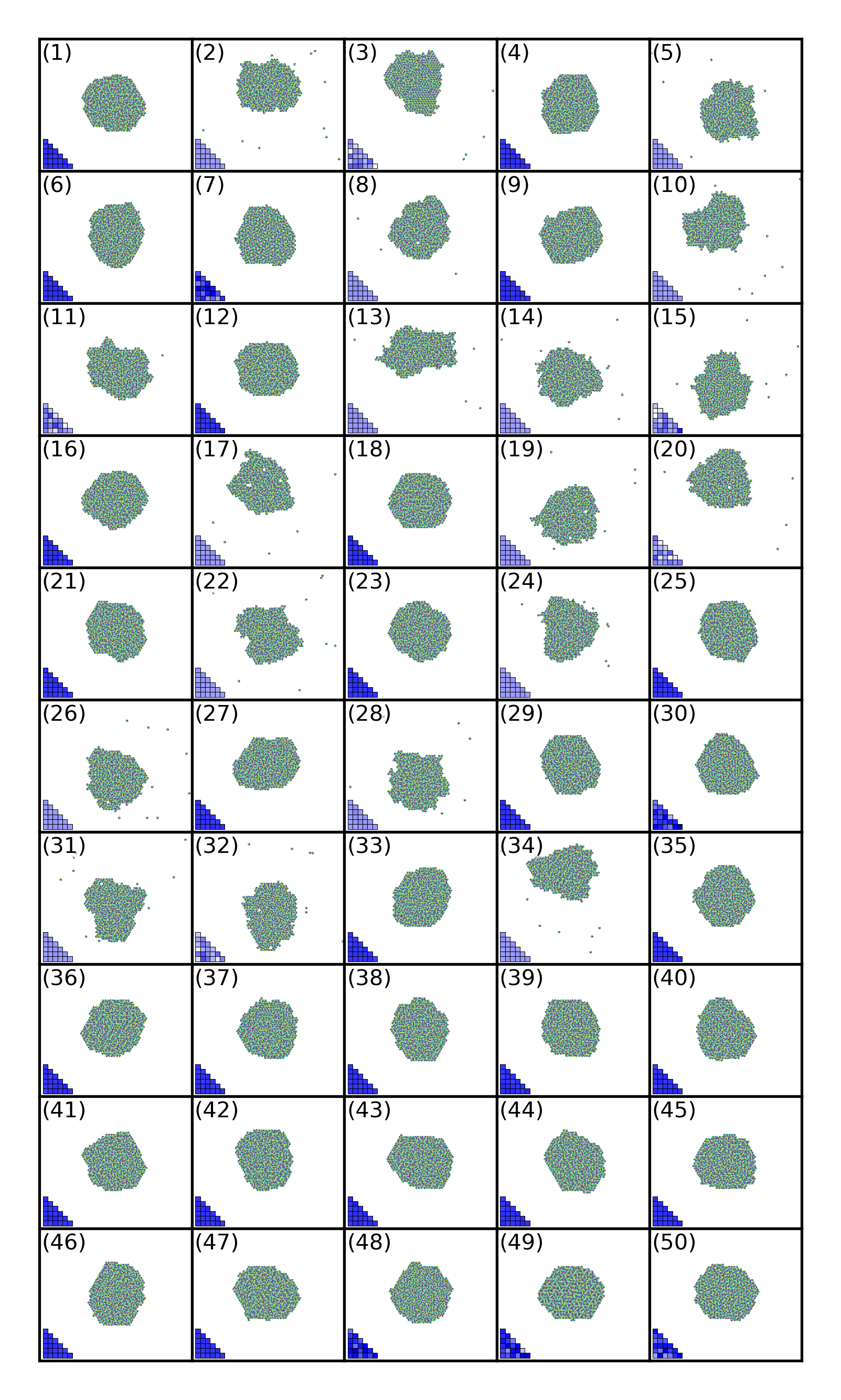}
    \caption{We label as liquids the dense aggregates that contain all or almost all particles in the simulation but do not display aggregate-wide orientational order}.
    \label{fig:label_liquid}
\end{figure}

In~\cref{fig:label_monomer,fig:label_oligomer,fig:label_fiber,fig:label_gel,fig:label_sponge,fig:label_crystal,fig:label_polycrystals,fig:label_liquid}, we show $50$ examples of manually labeled aggregates per category.  The visual criteria we use to distinguish between categories are the presence of interactions (monomers do not have interactions, as opposed to all the other categories), the dimensionality (monomers and oligomers are 0D, fibers are 1D, and gel, sponge, crystals, polycrystals, and liquids are 2D), the presence of orientational order (crystals and sponges
display orientational order, liquids do not, polycrystals and gels display only short-range orientational order), and the porosity (sponges are porous and crystals are not).


Some examples are not trivial to classify. Here, we illustrate our criteria by discussing some borderline cases. We label examples (6), (10) or (23) of~\cref{fig:label_monomer} as monomers and not oligomers because despite the presence of a few oligomers in the system, they do not always involve the same interactions, and a large fraction of the particles are unbound. 
We label examples (8), (32) or (49) of~\cref{fig:label_gel} as gels and not fibers because despite the one dimensional organization of the particles, it is not persistent enough to prevent those fibers to form loops. 
We label examples (19), (26) and (43) of~\cref{fig:label_gel} as gels and not sponges, because the crystalline organization is not systematically observed among the aggregates, and because these aggregate do not minimize the amount of surface particles by adopting a spherical shape. 

We label as crystals only the aggregates that are monocrystals. We include in the crystal category the aggregates displaying disclination lines~\cref{fig:label_crystal}(17,24,34) or point defects (16).
Conversely, polycrystals exhibits several competing orientational orders \cref{fig:label_polycrystals}(25,28), or short-range orientational order (20,24,48).
The liquids on the contrary, do not exhibit any visible orientational order~\cref{fig:label_liquid}. 

\begin{figure}
    \centering
    \includegraphics[width=0.45\linewidth]{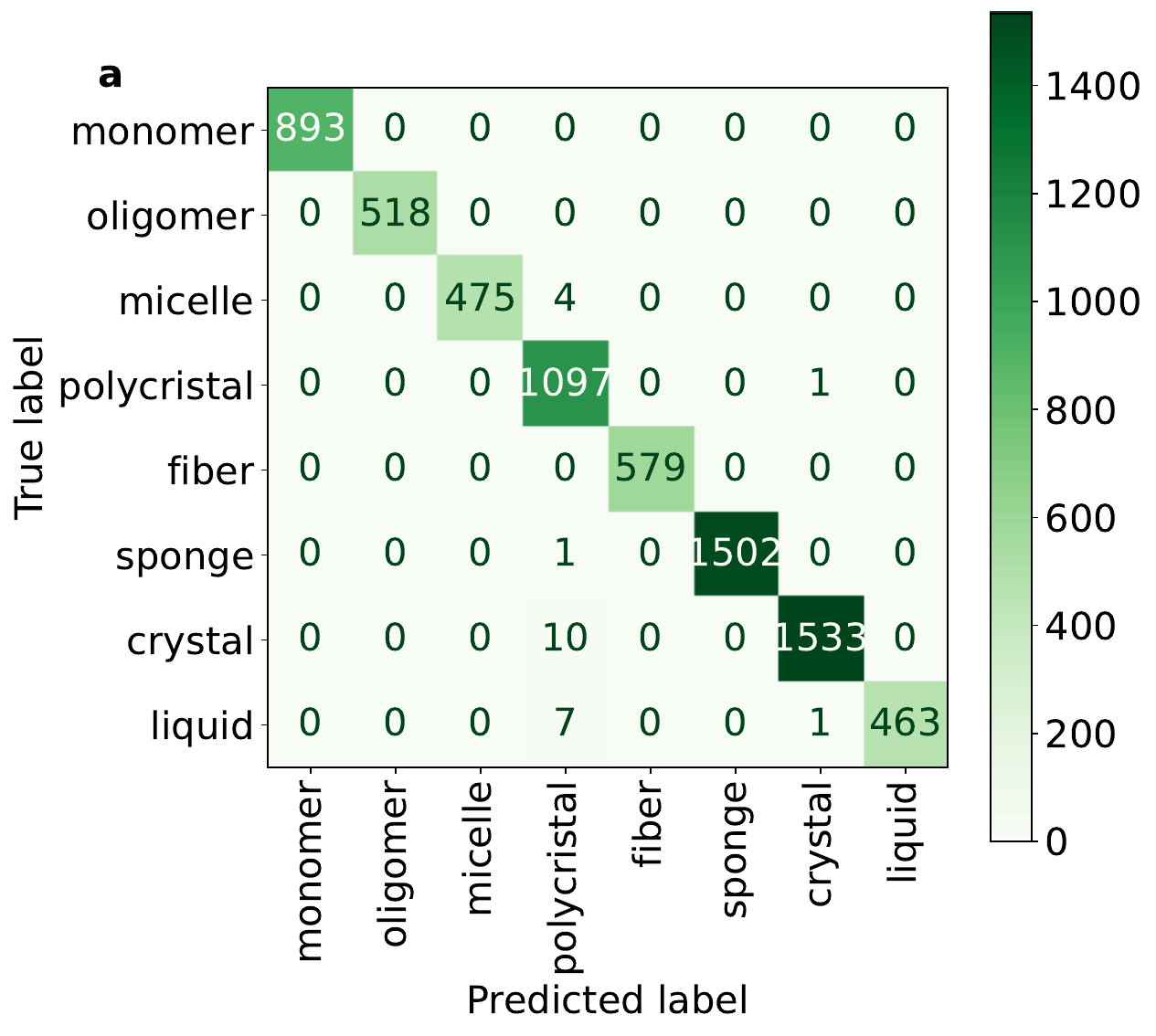}
    \includegraphics[width=0.45\linewidth]{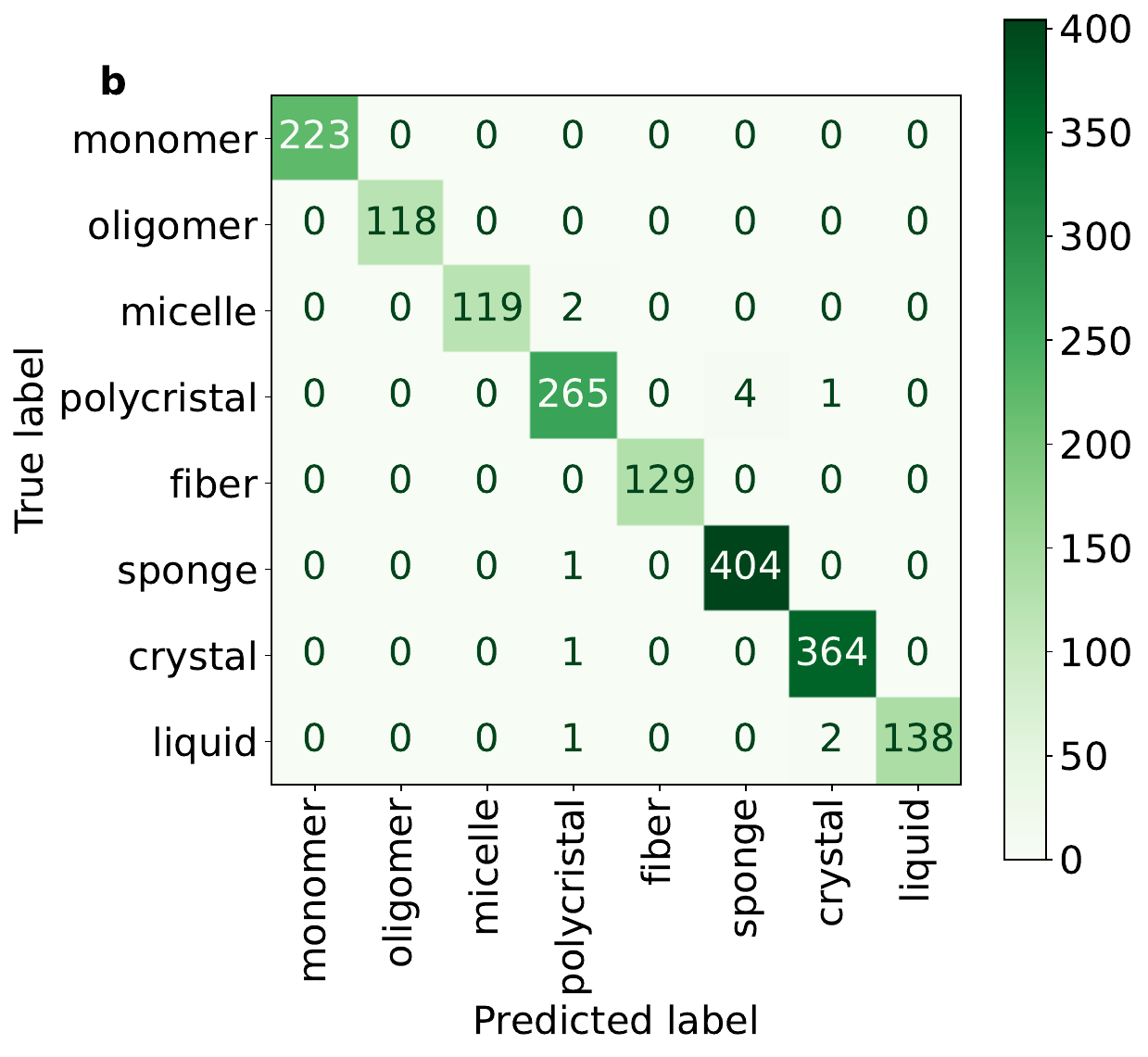}
    \caption{\textbf{The neural-network correctly classifies the aggregates into our eight categories.} We show the confusion matrix $C_{ij}$ (number of systems classified into category $j$ by the neural-network while hand-labeled as belonging to category $i$) for the training set (a) and the test set (b). The entries of the two matrices sum to $738\times12$, the number of labeled systems.}
    \label{fig:ML}
\end{figure}

\cref{fig:highest_scores} presents evidence that the neural-network learns our eight categories well. Specifically, it shows that for each system, the neural-network-computed probability of the most likely category is close to unity. In~\cref{fig:ML} we additionally show that there are very few misclassifications, both on the training and the test set, emphasizing that the characteristics used in our manual labeling are well learned by the neural-network. 

\begin{figure}
    \centering
    \includegraphics[width=\linewidth]{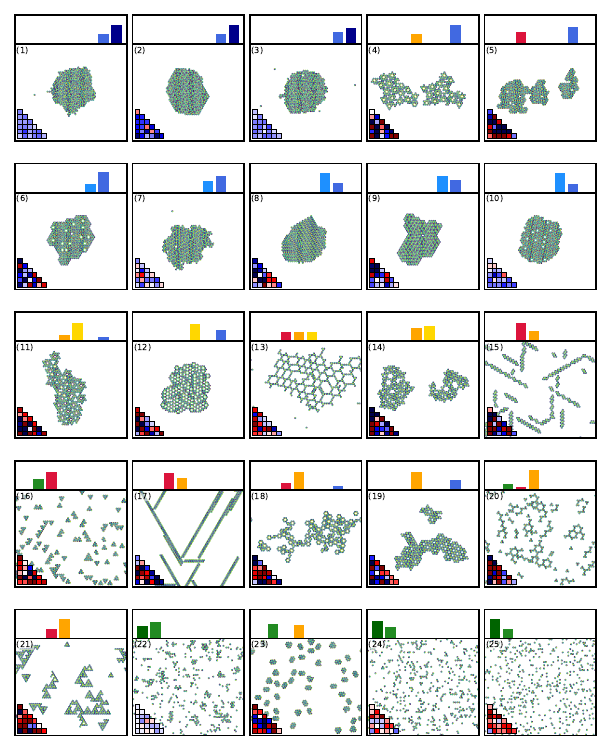}
    \caption{\textbf{Aggregates that have an ambiguous classification are in between two categories}. We show an image of the equilibrium configuration, the interaction map on the bottom left, and the probability vector outputted by the neural-network on top. The limits of the plots are between $0$ and $1$, the categories are ordered as in the rest of the paper, and the colors are as in \cref{fig:phase_diagram}}
    \label{fig:low_scores}
\end{figure}

In~\cref{fig:low_scores}, we also show some of the few examples for which the neural-network categorization is ambiguous, \textit{i.e.}, for which the prediction score defined in the main text is not close to unity. It concerns aggregates that have properties associated with two categories.
Example~(8) is a crystal, but has several disclination lines locally similar to a fiber. Example~(11) is a sponge, but does has the non-spherical shape similar to the gels. Particles in example~(16) assemble into trimers, and those trimers assemble in one dimensional structures, such that the aggregate is classified as a fiber. Example~(25) is a mixture of trimers and monomers.  Those examples of misclassification indicate that there are no entirely new aggregates morphologies that do not enter any of our eight categories.

\newpage
\subsection{Phase diagram}

\begin{figure}
    \centering
    \includegraphics[width=0.5\textwidth]{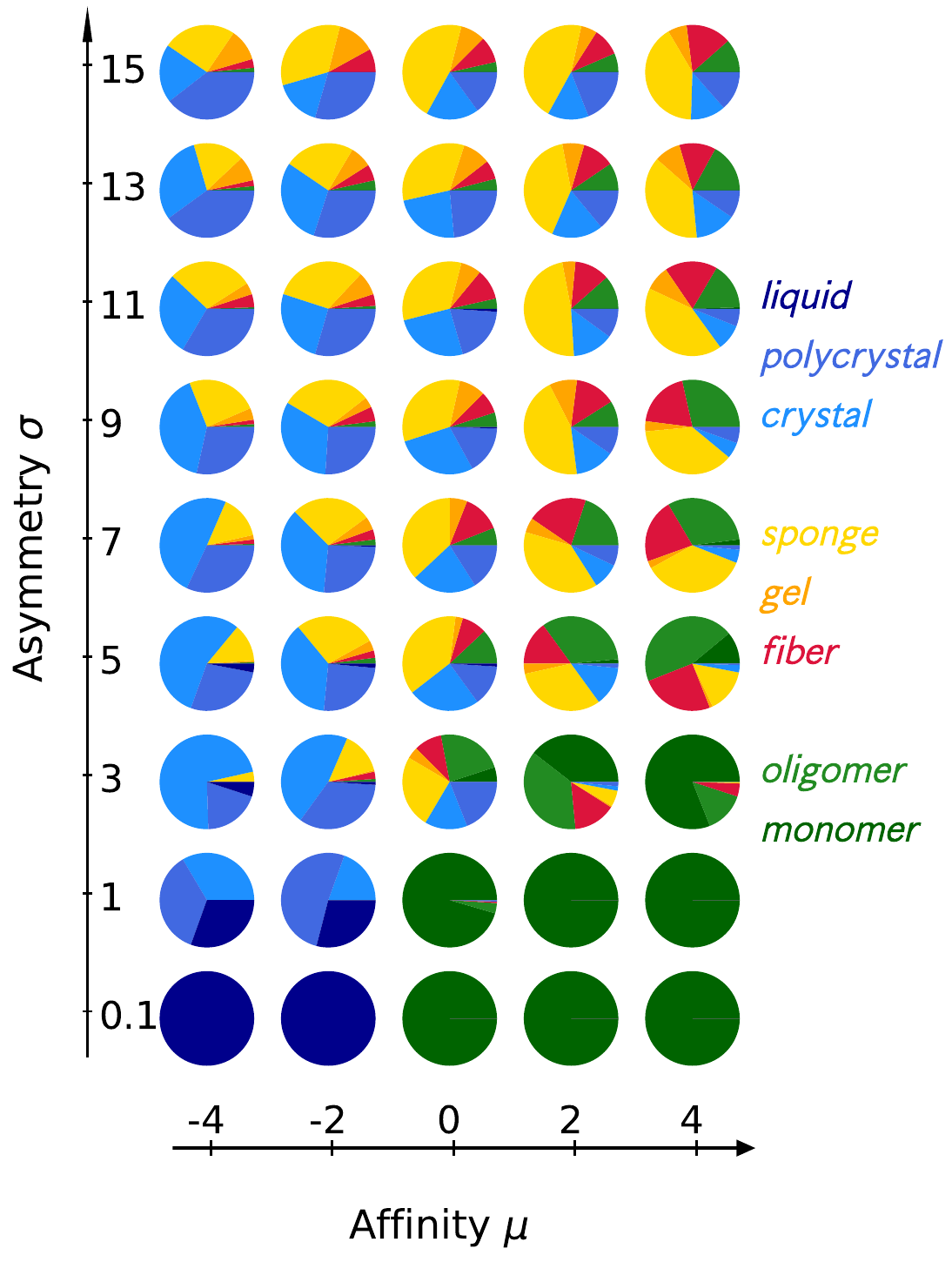}
    \caption{\textbf{The phase diagram of \cref{fig:phase_diagram} is not substantially modified when plotted as a function of the value of affinity and anisotropy of the distribution used to generate the interaction maps.}}
    \label{fig:phase_diagram_drawn_binning}
\end{figure}

\cref{fig:phase_diagram} shows data binned according to the measured average and standard deviation of each interaction map. In~\cref{fig:phase_diagram_drawn_binning} we show the same graph with the data binned according to the affinity $\mu$ and anisotropy $\sigma$ of the probability distribution of \cref{eq:proba} used to generate the interaction map. Both phase diagrams have the same tendencies: Interaction maps with small asymmetries and large affinities mostly form oligomer and monomers,
those with small asymmetries and small affinities form two-dimensional aggregates. Finally, interaction maps with large asymmetries and large affinities form diverse aggregate morphologies. This suggests that our phase diagram is robust to details in the binning procedure of its coordinates.

\subsection{The entropy of the particle's local surrounding discriminates between the aggregates of infinite size}
\label{subsec:entropy_local_structures}

We characterize the orientational order of the aggregates by measuring the entropy associated to the local environment of each particles. Since we run simulations of dilute systems, this methods is more appropriate than measuring a structure factor, the interpretation of which would be rendered difficult by the effect of the aggregate boundaries. Here, we define this entropy and show that it is generically low for sponges and crystals, intermediate for polycrystals, and large for liquids. 

\begin{figure}
    \centering
    \includegraphics[width=0.99\textwidth]{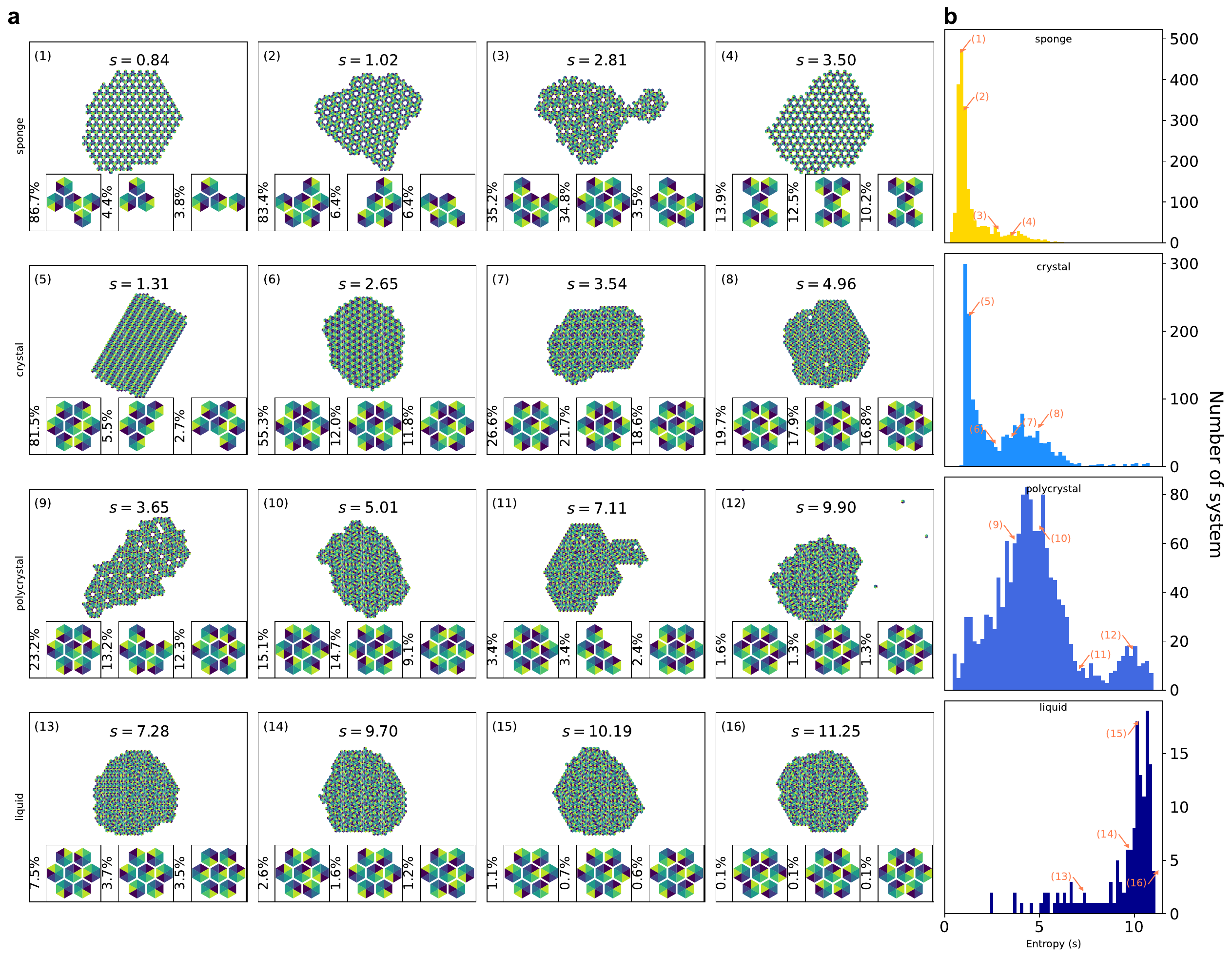}
    \caption{\textbf{A characterization of the local ordering of our aggregates through the entropy associated with a particle's local environment is highly consistent with our machine learning characterization}.
    (a)~For 16 systems among the categories sponge, crystal, polycrystal and liquid, we show a snapshot of the aggregate, the value of the entropy $s$ measured according to \cref{eq:entropy}, and the three most represented local environments in the inset, as well as the percentage of particles within this environment.
    (b)~The distribution of entropy values for aggregates in each of the mentioned categories shows that sponges and crystals have low entropy, polycrystals intermediate entropy, and liquids high entropy. }
    \label{fig:entropy_histograms}
\end{figure}

For each final aggregate structure produced by our simulations, we determine the local environment of each particle, \emph{i.e.}, the set of pair interactions it has with its 6 neighbors. If we fix the orientation of the particle in the center, there are 7 possible situations for each of its neighboring site: it can be a particle in one of the $6$ possible orientation, as well as an empty site. Therefore, there are $7^6$ possible local environments. We show examples of such local environments in the inset of \cref{fig:entropy_histograms}(a). The number of local environments observed in a system characterizes how ordered it is: in a crystal with one particle per unit cell, all the particles will have the same local environment, while in a high-temperature liquid where all the particles have fully random orientations, all possible local full-full environments are equally likely.

From the count of all the local environments observed within a system, we then measure the entropy of this system. We denote by $\alpha$ the index associated with a local environment, and by $p_\alpha$ the frequency with which this local environment is observed. The entropy then reads 
\begin{equation}
\label{eq:entropy}
    s = - \sum_{\alpha=1}^{7^6} p_\alpha \log_2(p_\alpha).
\end{equation}
The maximum value $s=\log_2(7^6)=16.8$ of this entropy is achieved when all local environments are equally likely. On the other end, if all the particles are in the same environment, the entropy is $s=0$. We show several typical examples of systems in \cref{fig:entropy_histograms}a. When the aggregates are perfectly ordered, such as in examples (1) or (5), one local environment dominates, while the other corresponds to the environment of the particle at the aggregate's boundary. 

We estimate the value of the entropy $s$ for all our simulation result, and find that it correlates with the aggregate category. \cref{fig:entropy_histograms}(b) shows the histograms of entropy values for the systems classified in all of our four categories corresponding to dense aggregates. We compute the entropy over $5$ snapshots of $500$ particles using \cref{eq:entropy}. While the sponges and crystals have entropy below $5$, the polycrystals have entropy values centered around $5$ and most liquids have entropy values above $10$. This measure also shows that we observe similar entropy values for crystals with a few defects, \emph{e.g.}, system (8) in the figure, and for polycrystals with two similar competing orientational orders [system (9)].

\subsection{Finding the best predictor of the aggregation category}
\label{subsec:other_interpretation}

\cref{fig:interpretationML} shows the learning accuracy of a few descriptors, that suggest that the propagability is an excellent descriptor of the aggregation category despite its relatively small size (it comprises 6 features). We have also considered many other, less effective descriptors, which we detail here. Each descriptor contains the average and standard deviation of the contact map in addition to the features discussed below.

\begin{figure}
    \centering
    \includegraphics[width=0.8\textwidth]{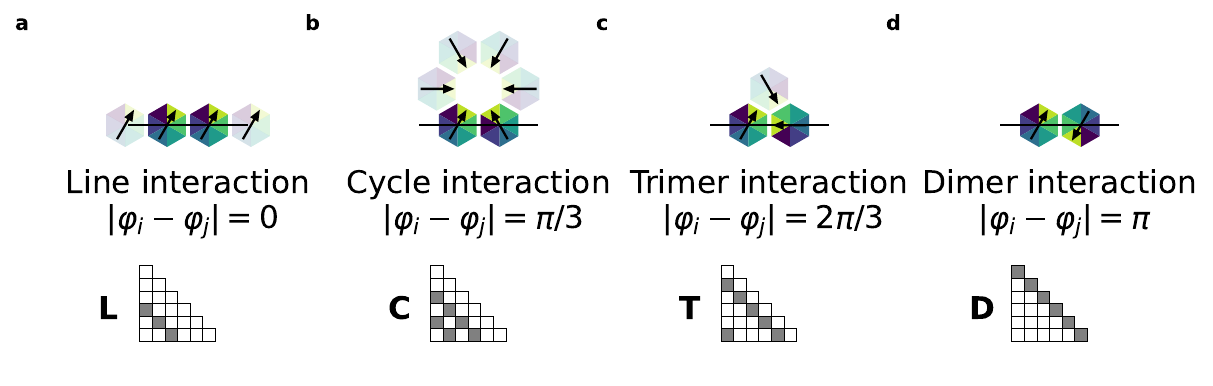}
    \caption{\textbf{The angle of an interaction characterize the typical motifs it leads to.} Depending on the angle of two neighboring particle orientations, the corresponding interaction leads to aggregate lines, trimer, cycle, or dimers. This angle corresponds to entries of the interaction map that are in the same diagonal (colored in gray).}
    \label{fig:angle}
\end{figure}

\begin{figure}
    \centering
    \includegraphics[width=0.8\textwidth]{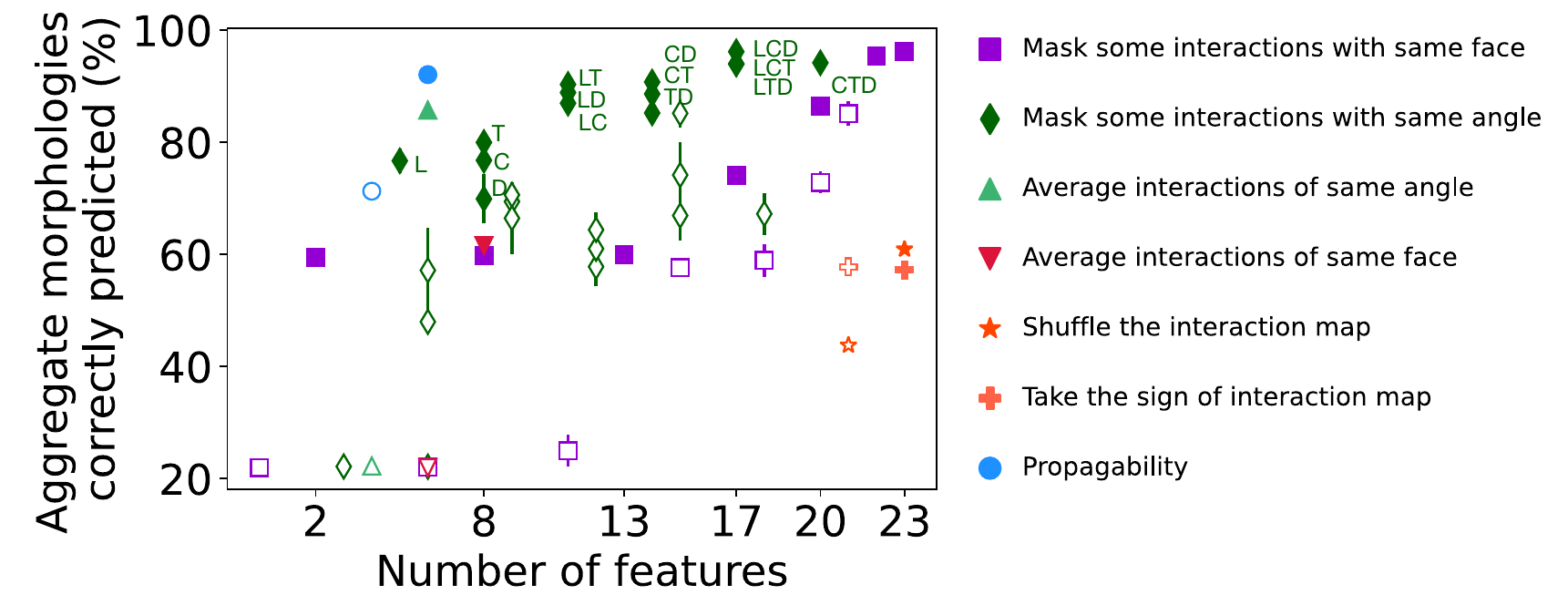}
    \caption{\textbf{The propagability’s high accuracy and small number of features makes it our best predictor.} For each test described in the text, here we show the accuracy of the prediction on the test set as a function of the number of features. We show the measured standard deviation for each descriptor. The letters next to the green symbols refer to the type of interactions that are not masked in the nomenclature of~\cref{fig:angle} (“L”=Line, etc). The empty symbols correspond to their equivalent full symbol, but the learning was performed without av($J$) and std($J$), and the number of features is $2$ less.}
    \label{fig:interpretationML_supp}
    
\end{figure}

Our first alternative descriptor is based on a similar idea as that depicted by the purple squares of \cref{fig:interpretationML}. These symbols correspond to descriptors comprised of a partially masked interaction map. While in that example we masked the rightmost columns of the matrix representation of the interaction map, \emph{i.e.}, all interactions corresponding pertaining to a subset of the faces of the particle, we may choose to mask the interactions corresponding to a specific \textit{angle of interaction}. This idea and the corresponding masked matrix elements are illustrated in \cref{fig:angle}. The resulting prediction accuracies are illustrated by dark green diamonds in \cref{fig:interpretationML_supp}. Overall, this methodology outperforms the masking baseline of the main text, and combinations including the line interactions are the most effective among the descriptors of this class.

Instead of simply masking some of the information contained in the interaction map, we also assess descriptors computed from its full specification, similar to propagability.
We first use the six values of the averaged face interaction, \textit{i.e.}, $\frac{1}{6}\sum_b J_{ab}$ for $a\in\llbracket 1,6\rrbracket$.
The resulting accuracy is indicated by the red downward facing triangle in \cref{fig:interpretationML}, and falls almost exactly on the purple baseline.
We next use the four average of the ``angle interaction'' categories defined in \cref{fig:angle}, which performs almost as well as the propagability (light green triangle in \cref{fig:interpretationML}). 

In the main text, we show that knowing the sign of the interaction is not sufficient to predict the aggregate morphologies. This implies that their strengths are crucial for this purpose. Conversely, here we ask whether knowing only the unordered list of interaction strengths enables good predictive power. We thus randomly shuffle the entries of each interaction map, leading to the orange star in~\cref{fig:interpretationML_supp}. This predictor performs about as badly as the signs-only predictor, highlighting the importance of the particles' geometry.

Finally, we compare the learning performances when av($J$) and std($J$) are not given to the neural network. Even when the whole $J$ matrix is given by the learning, the performances are increased of about $10\%$ by directly giving the values of av($J$) and std($J$) to the algorithm, revealing that those combinations are important to determine the aggregate category (as we already understood from \cref{fig:phase_diagram}), and are not trivial to learn (specifically because std($J$)) is a non-linear combination of the values of $J$. Even in the absence of av($J$) and std($J$), the propagability (empty blue circle) clearly outperforms the other measure, confirming that it captures the important physical properties of the interaction map.

\end{document}